\documentclass[reprint,
 amsmath,amssymb,
aps,
pra,
floatfix,
]{revtex4-1}

\usepackage{graphicx}
\usepackage{dcolumn}
\usepackage{bm}
\usepackage[caption = false]{subfig}
\usepackage{natbib}
\usepackage{url}
\usepackage{hyperref}
\usepackage[mathlines]{lineno}

\begin{document}

\title{Intrinsic Retrieval Efficiency for Quantum Memory: A Three Dimensional Theory
of Light Interaction with an Atomic Ensemble}

\author{Tanvi P Gujarati}
\author{Yukai Wu}
\author{Luming Duan}
 \email{lmduan@umich.edu}
 \altaffiliation[Also at ]{Center for Quantum Information, IIIS, Tsinghua University, Beijing 100084, PR China.}
 
\affiliation{%
 Department of Physics, University of Michigan, Ann Arbor, Michigan 48109, USA\\
}%

\date{\today}

\begin{abstract}
Duan-Lukin-Cirac-Zoller (DLCZ) quantum repeater protocol, which was proposed
to realize long distance quantum communication, requires usage of quantum
memories. Atomic ensembles interacting with optical beams based on
off-resonant Raman scattering serve as convenient on-demand quantum memories.
Here, a complete free space, three-dimensional theory of the associated read and write process for this quantum memory is worked out with the aim of
understanding intrinsic retrieval efficiency. We develop a formalism to
calculate the transverse mode structure for the signal and the idler photons
and use the formalism to study the intrinsic retrieval efficiency under
various configurations. The effects of atomic density fluctuations and atomic
motion are incorporated by numerically simulating this system for a range of
realistic experimental parameters. We obtain results that describe the
variation in the intrinsic retrieval efficiency as a function of the memory
storage time for skewed beam configuration at a finite temperature, which
provides valuable information for optimization of the retrieval efficiency in experiments.  

\begin{description}
\item[PACS numbers]
42.50.Ct, 03.67.-a


\end{description}
\end{abstract}

\maketitle


\section{\label{sec:level1}Introduction}

\begin{figure}
  \includegraphics[width=\linewidth]{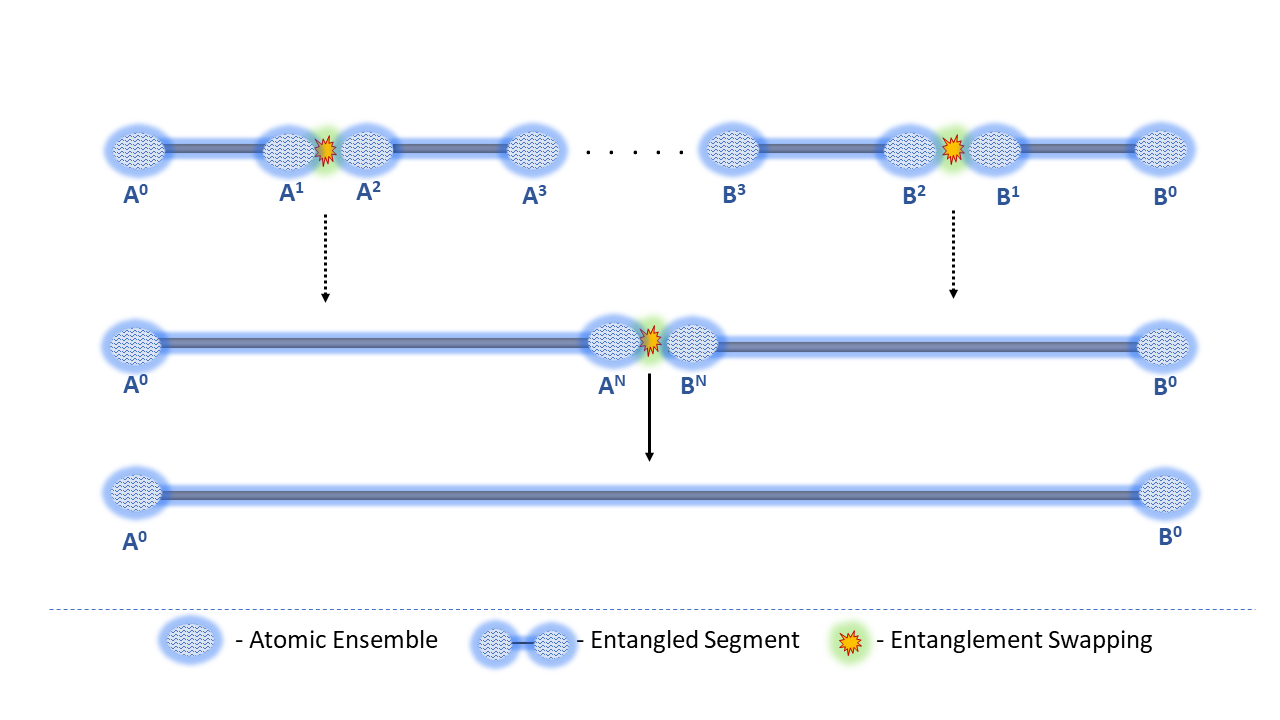}
  \caption{(color online) Working of quantum repeaters. To generate entanglement between two distant nodes $A^{0}$ and $B^{0}$, we start by dividing the total distance into smaller segments $A^{0}-A^{1}$,$A^{2}-A^{3}$,...,$B^{3}-B^{2}$,$B^{1}-B^{0}$ with their corresponding nodes. Entanglement is first generated in the smaller segments between each of these nodes independently. By entanglement swapping between two neighbouring nodes, e.g. $A^{1}$ and $A^{2}$, the entanglement can be extended over a longer segment $A^{0}-A^{3}$. With every successful entanglement swapping step, the generated extended entanglement between two nodes on a segment must be purified. By successive swapping and purification in a hierarchical manner, entanglement can be generated over the original distance between $A^{0}$-$B^{0}$.}
  \label{fig:qtrep}
\end{figure}

Quantum communication relies on the ability of generating quantum entangled
states over large distances. One way to accomplish this goal is to create
entanglement between distant units with the help of appropriate communication
channels between them. Typical carriers of quantum information, the photons,
suffer from losses due to absorption and decoherence in the transfer channel.
This leads to an exponential decay of communication fidelity with increasing
distance of communication. The way out of this problem is to use quantum
repeaters \cite{Briegel98}. Quantum repeaters are modeled on the divide
and conquer approach. The entire length over which entanglement is to be
created is broken down into smaller segments. Physical systems at the ends of
each smaller segment can be efficiently entangled because of smaller lengths
between them [Fig.~(\ref{fig:qtrep})].
Then, entanglement can be generated between two adjacent segments
by entanglement swapping using neighboring systems (\cite{Bennett93}%
,\cite{Zukowski93}). This process can be repeated until entanglement is
generated over the full length. At each step though, the entanglement needs to
be purified which is a probabilistic process. Thus, to extend entanglement
over two adjacent segments one has to wait till entanglement is generated and
purified over each segment \cite{Duan01}. The upshot is that quantum
repeater protocols require quantum memories (\cite{Duan01},\cite{Sangouard11})
that can store the entanglement for one segment till it is created in the
neighboring segment.\\  

In 2001, the DLCZ quantum repeater scheme was introduced. This scheme showed a
way to generate heralded entanglement over a distance by using atomic
ensembles as individual memory units in combination with linear optics and
single-photon detectors. The atomic ensembles form the physical systems or
nodes at the end of each segment which can store de-localized spin-wave state
when entangled. These nodes are connected by fiber optic cables which serve as
the communication channels between ensembles allowing efficient transfer of
photons. Entanglement between two neighbouring nodes on adjacent segments can
be generated by converting the stored spin-waves in the atomic ensembles into
correlated photons and performing beam-splitter measurements on them. The
generation and detection of a single photon from the atomic ensemble, in the
absence of which way information, makes the two segments get entangled. Memory nodes
based on atomic ensembles as opposed to single atoms make strong coupling
between atoms and photons possible due to collective effects of a large number
of atoms. A brief description of the DLCZ scheme and the collective effects in
atomic ensembles is provided in Sec. \ref{sec:level2} for completeness.
Following the DLCZ scheme, many experiments have demonstrated remarkable
advances towards quantum repeaters (\cite{Pu2017},\cite{Chou05},\cite{Chou07}).

The atomic ensembles that act as individual nodes to store de-localized
quantum entangled states must satisfy a few important properties. They should
have long storage lifetimes and high retrieval efficiency \cite{Sangouard11}. Storage lifetimes of about milliseconds to seconds have been achieved in
quantum memories with atomic gases (\cite{RZhao09},\cite{Zhao09}%
,\cite{Yang16},\cite{Dudin13}). 
Intrinsic retrieval efficiency
(IRE) is defined as the probability of retrieving an idler photon in a
particular spatio-temporal mode from the stored spin-wave excitation in the
atomic ensemble conditioned on the successful detection of signal photon in
the write process. Detailed theoretical description of IRE is given in Sec.~\ref{sec:level3}. The spatio-temporal mode of the signal and the idler photon
must have a high overlap with single mode optical fibers which are used in
experiments to collect and propagate these photons for interference and
detection. In our definition of the intrinsic retrieval efficiency, we include
contributions from mode-overlap between emitted photon field and the optical
fiber field as it is an integrated part of photon read out process in
experiments. Because of the collective effects of atoms involved in the
light-matter interaction, the read-out photon is highly correlated with the
spin-wave excitation stored in the atomic ensemble. High IRE values are
extremely important for reasonable entanglement distribution rates
(\cite{Duan01,Sangouard11}). For example, as is stated in
\cite{Sangouard11}, 1\% reduction in IRE, from 90\% to 89\%, increases the
entanglement distribution time over a distance of 600Km by 10\%-14\% for the DLCZ protocol and its
variants. Calculations in \cite{Duan01} show that the scaling of the total
time of entanglement generation between two distant atomic ensembles with the
number of repeater nodes critically depends on the IRE. Free space IRE in
experiments with cold atom ensembles is at best about 50\% \cite{Laurat06}.
For atomic ensembles confined to cavities, IRE of more than 70\% has been
achieved (\cite{Yang16},\cite{Simon07}). The IRE is sensitive to decoherence
due to stray magnetic fields, atom loss as well as dephasing of the spin-wave
caused by atomic motion. To understand the exact nature of the IRE, it is
important to study the full three dimensional profile of the spin-wave
excitation stored in the atomic ensemble and how it gets mapped into the
transverse (angular) profile of the emitted photon following the read-out
process. Our goal in this paper is to understand the intrinsic memory
retrieval efficiency by performing a thorough three-dimensional quantum
mechanical calculation that also takes into account the mode matching between
the emitted photons and single photon collection fibers.\\

We would like to note that previous efforts to theoretically describe the
read-write process using the Maxwell-Bloch formalism work with one dimensional
description of the atomic density and electric field propagation
\cite{Gorshkov07}. Such a description works well only when we assume that the
write beam waist is much broader than the beam waist of the emitted
photon. Recent experiments \cite{Pu2017} use beam parameters which are
marginally close to not being described by this theoretical treatment. The transverse mode profile of the electric fields play an important role for understanding IRE. As we shall show in our results, IRE is sensitive to the ratio of the beam
waists between the write and signal/idler photon beams. It is also important to
note that the Maxwell-Bloch approach doesn't describe the electric field that
gets scattered from the atoms. This scattered field is what we are
interested in when calculating IRE as the desired spatio-temporal mode of the
emitted photon continuously changes to the other scattered modes which
contribute to noise. One of the ways of improving the IRE is by increasing the
optical depth. This can be achieved by taking longer atomic samples in the
direction of light propagation without increasing the overall atomic density.
For longer geometries of atomic samples it becomes essential to look at the
variation of the transverse profile of the light beams due to diffraction.\\

A three-dimensional formalism for calculating the field modes of light
scattered from an ensemble of hot atomic gas was presented \cite{Duan02}. In
this calculation, the atomic positions were averaged over the duration of
interaction with light to get the emitted photon mode profile. This averaging
significantly simplifies the calculations to get the mode profile of the
photon correlated with the symmetric collective spin wave state. Since, we are
interested in describing cold atomic ensembles, such averaging over positions
cannot be done. One of the interesting results from this calculation
in \cite{Duan02} suggested that atomic density fluctuations give rise to
intrinsic mode mismatching errors. We find that atomic density
fluctuations have a significant role to play when determining IRE.\\ 

This paper is organized as follows: in Sec. \ref{sec:level2} the interaction
scheme between the atomic ensemble and light is discussed with the aim of
understanding the IRE of a quantum memory unit based on such an interaction. In
Sec. \ref{sec:level3} a detailed theoretical analysis for the write and read
process defining the storage and retrieval of quantum spin wave is presented.
Sec. \ref{sec:level6} focuses on the results obtained by numerical simulations
of atoms in a node subject to motion. In the final Sec. \ref{sec:level7} we
revisit the results and conclude the discussion.

\section{\label{sec:level2}Read and write process of an atomic quantum memory}

In this section, we will take a close look at the DLCZ scheme and define the
associated atoms-light interaction configuration.\\

\begin{figure}
  \includegraphics[width=\linewidth]{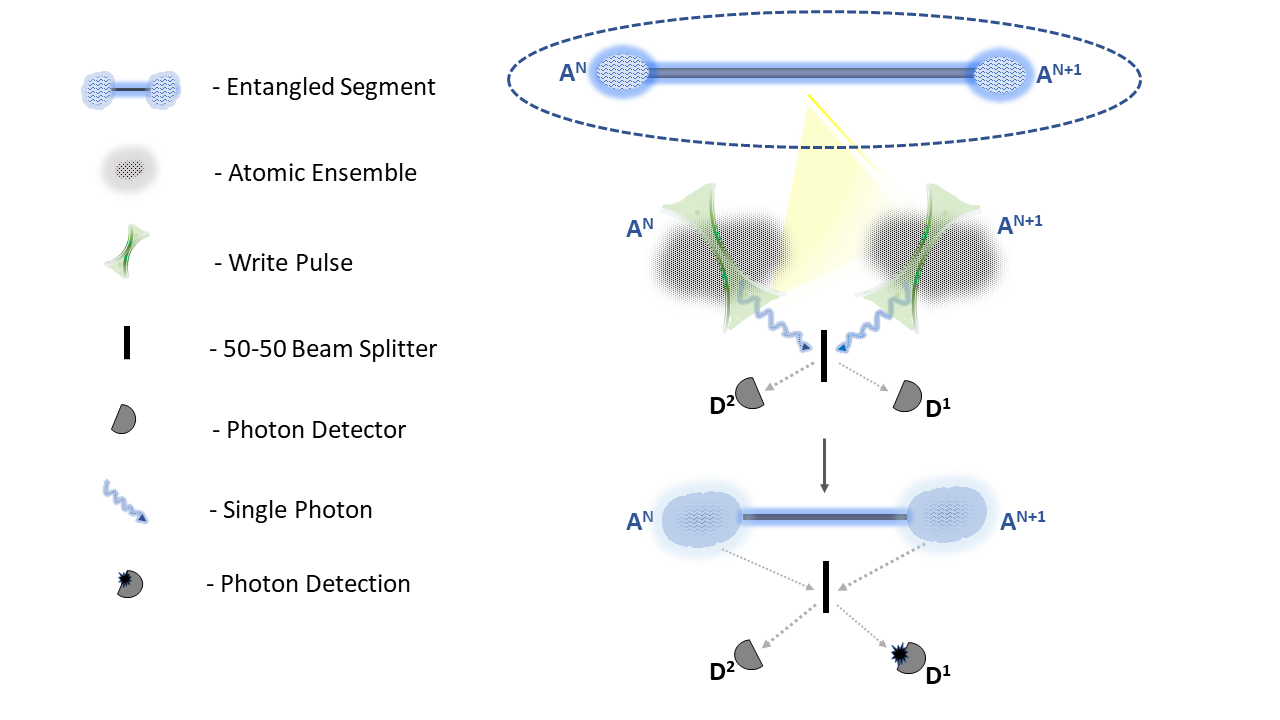}
  \caption{(color online). Entanglement generation between two atomic ensembles $A^{N}$ and $A^{N+1}$. The atomic ensembles to be entangled are simultaneously excited with weak off-resonant Raman pulses, called the write pulse. A photon corresponding to the atomic spin wave mode emitted from any one of the ensembles is sent through the 50-50 beam-splitter. The output arms of the beam-splitter are in-turn coupled to single photon detectors. For ideal photon detectors, a click in any of the two detectors, e.g. $D^{1}$ in this case, heralds the generation of entanglement between the two atomic ensembles $A^{N}$ and $A^{N+1}$.}
  \label{fig:entgen}
\end{figure}

As shown in Fig.~(\ref{fig:qtrep}), to generate entanglement over $A^{0}$ and
$B^{0}$, we split the intermediate distance into multiple smaller segments and
perform entanglement generation for each segment followed by entanglement
swapping between neighbouring segments sequentially. Let us look at the
entanglement generation step first. A pictorial representation of the setup
for entanglement generation between two atomic ensembles on a segment is shown
in Fig.~(\ref{fig:entgen}). The two ensembles $A^{N}$ and $A^{N+1}$ are
simultaneously excited with weak Raman pulses (write pulse), such that there
is a small but definite probability of one of the ensembles emitting a photon
correlated with the coherent spin-wave mode in the atomic ensemble
\cite{Duan01}. The photon generated from either of the samples is coupled to
optical fibers and made to interfere at a 50-50 beam-splitter coupled to
single photon detectors at the output arms. If either of the detectors clicks,
that heralds entanglement between the two ensembles. This is how entanglement
is generated within each segment of the quantum repeater scheme.\\

\begin{figure}
  \includegraphics[width=\linewidth]{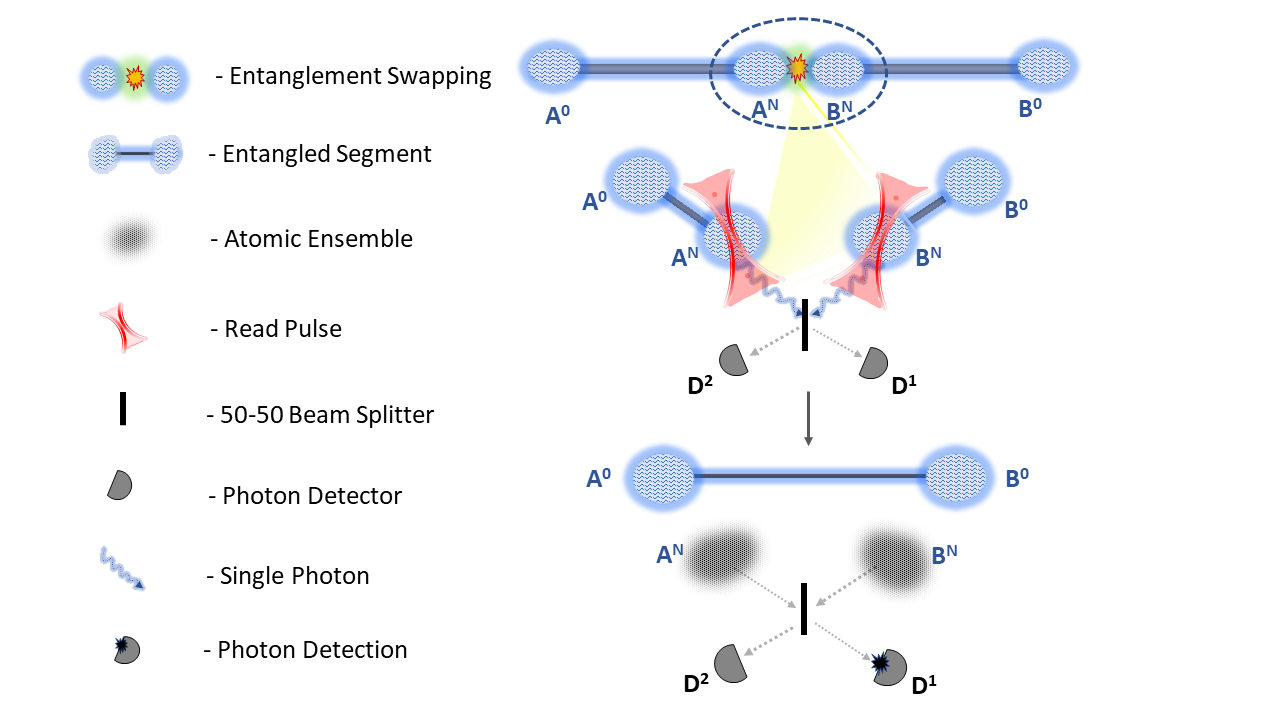}
  \caption{(color online).Entanglement Swapping between two neighbouring entangled segments. Given two entangled segments $A^{0}-A^{N}$ and $B^{N}-B^{0}$, entanglement is generated between atomic ensembles $A^{0}-B^{0} $ by entanglement swapping between ensemble $A^{N}-B^{N}$. The atomic spin-wave modes in the neighbouring ensembles $A^{N}$ and $B^{N}$ are converted into photons using a strong and broad read-out pulse. Photons emitted by the atomic ensembles are coupled to a 50-50 beam-splitter. The output from the beam-splitter is coupled to single photon detectors. Whenever one of the detectors registers a photon, the atomic ensembles $A^{0}-B^{0}$ get entangled due to entanglement swapping.}
  \label{fig:entswap}
\end{figure}

Once we have two such adjacent entangled segments eg. $A^{0}-A^{N}$ and
$B^{N}-B^{0}$ in Fig.~(\ref{fig:entswap}), we can carry out the next step of
entanglement swapping as follows. The ensembles $A^{N}$ and $B^{N}$ are
simultaneously excited with strong read-out pulses, such that there is a high
probability of a stored spin-wave atomic excitation getting converted into a
highly directional photon. These photons are collected and made to interfere
at another 50-50 beam-splitter connected also to single photon detectors. If
there is a click in either of the detector arms, that would lead to
entanglement of the ensembles $A^{0}-B^{0}$. The necessary requirement as
discussed previously is that the entanglement in either segment needs to be
stored until entanglement in the other segment can be generated and purified.
The process of entanglement generation, purification and swapping can now be
repeated to create entanglement sequentially between ensembles farther and
farther apart. The details of read and write process for each atomic ensemble
are given below.\\ 

\begin{figure}
  \includegraphics[width=\linewidth]{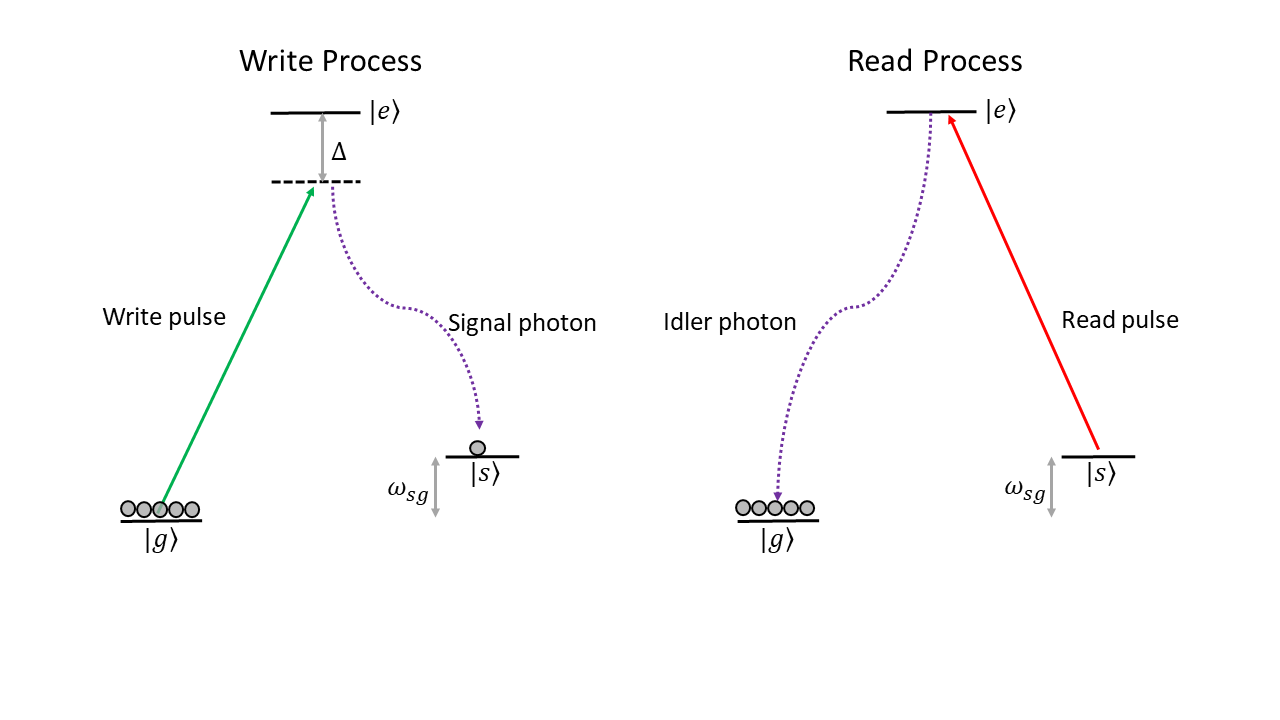}
  \caption{(color online). Atomic level diagram for the DLCZ protocol. Every atom in the atomic ensemble is considered to have a three level $\Lambda$ structure. Levels $|g\rangle$ and $|s\rangle$ are two metastable states separated by frequency equal to $\omega_{sg}$, with forbidden dipole transition between them. Level $|e\rangle$ is an excited state. At $t = 0$ all atoms are in the ground state $|g\rangle$. In the write process, the atomic ensemble is excited with a classical write-pulse that is detuned from the $|g\rangle$-$|e\rangle$ transition by a frequency $\Delta$. With the emission and detection of a signal photon on the $|e\rangle$-$|s\rangle$ transition the write process is complete with one atom excited to the $|s\rangle$ level. In the read process, the ensemble is excited with a strong on resonance read-pulse for $|s\rangle$-$|e\rangle$ transition. The emission and detection of a highly directed idler photon from $|e\rangle$-$|g\rangle$ concludes the read process.}
 \label{fig:atmstr}
\end{figure}

Consider an atomic ensemble with $N_{a}$ atoms with a $\Lambda$ level
structure as shown in Fig.~(\ref{fig:atmstr}). There are two metastable ground
levels, $|g\rangle$ and $|s\rangle$ having long lifetimes and an excited level
$|e\rangle$. All atoms are initially prepared in the ground state $|g\rangle$.
The atoms in the ensemble are acted upon with a weak off-resonant laser pulse,
the write-beam, on the $|e\rangle$-$|g\rangle$ transition. With some small
probability a single photon, called the signal photon, corresponding to
$|e\rangle$-$|s\rangle$ transition gets emitted spontaneously.\\

This is the two-photon Raman scattering process that results in a
transition of one atom from $|g\rangle$ to $|s\rangle$ level. Information of
which atom in the ensemble emitted the photon is lost for a far field
detection of the photon. This leads to the creation of a coherent collective
atomic spin-wave state of the form:
\begin{equation}
    |\Psi_{a}\rangle = \sum_{j = 1 }^{N_{a}} C_{j}e^{\textit{i}(\textbf{k}_{W}-\textbf{k}_{S})\cdot\textbf{r}_{j}}|g\rangle_{1}|g\rangle_{2}...|s\rangle_{j}....|g\rangle_{N_{a}}
\end{equation}
where $\mathbf{k}_{W}$ and $\mathbf{k}_{S}$ are the wavevectors associated
with the write-beam and the emitted signal photon respectively and
$\mathbf{r}_{j}$ is the position vector for $j^{th}$ atom. The complex
coefficients $C_{j}$ are dependent on the shape of the laser profiles at the
$j^{th}$ atomic position. With the detection of the signal photon, write
process is complete and information is now stored in the coherent atomic
spin-wave. \\

Now, we move on to describe the read-process. After a certain time $T_{m}$,
the storage time, a strong classical laser pulse (read pulse) resonant with
the $|e\rangle$-$|s\rangle$ transition is made to shine on the atomic ensemble
such that any atom in the $|s\rangle$ state gets excited to the $|e\rangle$
state. The atom in $|e\rangle$ state emits an idler photon to relax back to
the $|g\rangle$ state. The atomic quantum state after this process is
proportional to:
\begin{equation}
   \sum_{j=1}^{N_{a}}D_{j}e^{\textit{i}(\textbf{k}_{W}-\textbf{k}_{S})\cdot\textbf{r}_{j}}e^{\textit{i}(\textbf{k}_{R}-\textbf{k}_{I})\cdot\textbf{r}'_{j}}|g\rangle_{1}|g\rangle_{2}...|g\rangle_{N_{a}} 
   \label{eq:intf}
\end{equation}
where, $\mathbf{k}_{R}$ and $\mathbf{k}_{I}$ are the wavevectors corresponding
to the read beam and the emitted idler photon respectively. The position of
the $j^{th}$ atom after the storage time $T_{m}$ is given by $\mathbf{r}%
^{\prime}_{j}$. Because of finite temperatures of the atomic sample,
$\mathbf{r}_{j}$ is generally different from $\mathbf{r}^{\prime}_{j}$. For
the calculations henceforth, we assume that the atomic ensemble is a cold-atom
sample having a temperature of about 30$\mu K$ obtained by cooling a MOT
sample further via Polarization Gradient Cooling technique. The coefficients
$D_{j}$ are weights associated with the $j^{th}$ atom that depend on the
atomic positions $\mathbf{r}_{j}$ as well as $\mathbf{r}^{\prime}_{j}$ and
properties specific to the atom-light interaction like polarization, dipole
moment and beam parameters. Eq.~(\ref{eq:intf}) tells us that the amplitude of emission for the idler photon in the $\mathbf{k}_{I} $ direction is determined by interference between all the atoms of the ensemble scaled by factors
$D_{j}$. Because of constructive interference between all atom contributions, the idler photon is emitted in a well specified direction based on the phase matching condition. 
\begin{equation}
(\textbf{k}_{W}-\textbf{k}_{S})\cdot\textbf{r}_{j} + (\textbf{k}_{R}-\textbf{k}_{I})\cdot\textbf{r}'_{j} = 0
\label{eq:phasemat}
\end{equation}
We shall see from the calculations in the next sections, the intrinsic
retrieval efficiency is acutely affected by the interference condition. As
discussed in \cite{Sangouard11}, complete constructive interference is
possible only when the atoms don't move within the storage time ($\mathbf{k}%
_{W} + \mathbf{k}_{R} = \mathbf{k}_{S} + \mathbf{k}_{I}$) or when the beams
are colinear ($\mathbf{k}_{W} = \mathbf{k}_{S}$, $\mathbf{k}_{R}%
=\mathbf{k}_{I}$). In experiments with cold atomic gases, both these
conditions are seldom implementable. Because of position dependent weights
associated with the angular profile of the light and atomic spin-wave and
non-zero energy difference between the two ground levels, unit IRE cannot be
achieved.\\

With the basic idea of the protocol and importance of retrieval efficiency in
mind, let us now look at the full derivation of the mathematical expression of retrieval efficiency with a complete 3-D analysis.\\

\section{\label{sec:level3}Theoretical formulation of the intrinsic retrieval efficiency}
We will now formulate the interaction between light and
the atomic ensemble which acts as a temporary storage for quantum
entanglement. We shall also describe IRE formally and calculate it using the quantum theory
of light-matter interactions. As is already seen in Sec. \ref{sec:level1}, the
intrinsic retrieval efficiency is defined as the probability of getting the
desired photon from the stored atomic spin-wave. \\

After interacting with the write beam, the quantum state of the atomic ensemble and the emitted photon is expressed as:
\begin{equation}
    |\Psi\rangle_{W} = \int \frac{d^{3}\textbf{k}}{(2\pi)^{3}}\sum_{j=1}^{N_{a}} C_{j}(\textbf{k})|s\rangle_{j}|\textbf{k}\rangle_{ph} 
    \label{eq:writestate}
\end{equation}
where:
\begin{equation}
|s\rangle_{j} \leftarrow |g\rangle_{1}|g\rangle_{2}...|s\rangle_{j}...|g\rangle_{N_{a}}
\label{eq:state-s}
\end{equation}
and $C_{j}(\mathbf{k})$ is the photon wave function given an atomic excitation
for atom $j$ . Sum over $j$ adds contribution of all the atoms of the sample
and the integration over $\mathbf{k}$ for all the wave-vectors.\\

Detection of the write photon can be expressed as the overlap of the above
state in Eq.~(\ref{eq:writestate}) with a transverse Gaussian electric field mode coupled to the single mode optical fiber. The resulting quantum state after this overlap is the
obtained spin-wave state. This state after appropriate normalization gives the
initial condition of the atomic ensemble for the read process. After the read
process, the resulting photon quantum state can be described as:
\begin{equation}
|\Psi\rangle_{R} = \int \frac{d^{3}\textbf{k}}{(2\pi)^{3}}\sum_{j=1}^{N_{a}} D_{j}(\textbf{k})|\textbf{k}\rangle
\end{equation}
while the atomic ensemble is back to its ground state $|g\rangle^{\otimes
N}=|g\rangle_{1}|g\rangle_{2}...|g\rangle_{N}$. We can again represent
detection of the emitted read photon as an overlap of the emitted photon state
with transverse Gaussian field. The squared norm of this overlap would
correspond to the desired IRE. In the following subsections, we shall derive
the explicit expression of this quantity.

\subsection{\label{sec:level4}The Write Process}
For the atomic level structure given in Fig.~(\ref{fig:atmstr}), in the write
process, the atomic ensemble is excited by a weak and short off-resonant Raman
pulse (the write pulse) coupled to the $|g\rangle$-$|e\rangle$ transition. We
treat this interaction semi-classically, by taking classical light pulse interacting with a quantum atomic system. The electric field associated with the write pulse is given as:
\begin{equation}
    \textbf{E}^{w}(\textbf{r},t)=\frac{1}{2}\left[  \hat{\epsilon}^{w}E^{w}(\textbf{r},t)e^{\textit{i}(\textbf{k}^{w}\cdot\textbf{r}-\omega^{w}t)}+\text{c.c.}\right]
\end{equation}
Where $\omega^{w}=k^{w}c$ is the carrier frequency of the write pulse and
$|\mathbf{k}^{w}|=k^{w}$. Also $\hat{\epsilon}^{w}$ is the unit direction of the field. It is assumed to be a square pulse of width $T_{w}$ time
units.\\

The spontaneously emitted photon corresponding to the $|e\rangle$-$|s\rangle$ transition is treated quantum mechanically. The electric field associated with the emitted signal photon is described by the sum of all the free field modes:
\begin{equation}
\hat{\textbf{E}}(\textbf{r})=\sum_{\tau}\int \frac{d^3\textbf{k}}{(2\pi)^{3}}\left[\hat{\epsilon}_{\textbf{k},\tau}f(k)e^{\textit{i}\textbf{k}\cdot\textbf{r}}a_{\textbf{k},\tau}+\text{h.c.}\right]
\end{equation}
In the above expression, $\mathbf{k}$ stands for the wavevector of the emitted
photon and $\tau$ for one of the two independent polarization directions given
a wavevector. The operators $a_{\mathbf{k},\tau}$ and its Hermitian conjugate
$a^{\dagger}_{\mathbf{k},\tau}$ are the annihilation and creation operators
for the given wavevector $\mathbf{k}$ and polarization $\tau$. The dispersion
relation is given as $\omega_{k}=|\mathbf{k}|c$. Also for free space normal
modes, the expression for the mode function $f(k)$ is:
\begin{equation}
f(k)=\textit{i}\sqrt{\frac{\hbar \omega_{k}}{2\varepsilon_{0}}}
\end{equation}
where $\varepsilon_{0}$ is the free space permittivity. Throughout this paper
we set $\hbar=1$ for simplicity.\\

We assume that there is no atom-atom interaction in the system. The atom-field
interaction Hamiltonian taken here is the dipole interaction with minimal
coupling. Under the rotating wave approximation (RWA) we get the following
Hamiltonian given in Eq.~(\ref{eq:wrHam}). Note that spontaneous emission from
the state $|e\rangle$ to $|g\rangle$ is ignored as it is not important for our
purpose. Taking the energy of the $|g\rangle$ state, $\omega_{g}$, to be our 0 reference, the write Hamiltonian is then:
\begin{eqnarray}
H^{w}&=&\sum_{\tau}\int \frac{d^{3}\textbf{k}}{(2\pi)^{3}}~\omega_{k}a^{\dagger}_{\textbf{k},\tau}a_{\textbf{k},\tau}+\sum_{j=1}^{N_{a}} (\omega_{eg}\sigma_{ee}^{j}+\omega_{sg}\sigma_{ss}^{j})\nonumber\\
&+&\sum_{j=1}^{N_{a}}\bigg[\Omega_{eg,j}^{w}e^{\textit{i}(\textbf{k}^{w}\cdot\textbf{r}_{j}-\omega^{w}t)}\sigma_{eg}^{j}\nonumber\\
&+&\sum_{\tau}\int \frac{d^{3}\textbf{k}}{(2\pi)^{3}}~g_{es,\tau}(\textbf{k})e^{\textit{i}\textbf{k}\cdot\textbf{r}_{j}}\sigma_{es}^{j}a_{\textbf{k},\tau}+\text{h.c.}\bigg]
\label{eq:wrHam}
\end{eqnarray}
where:
\begin{eqnarray}
\omega_{ab} &=& \omega_{a}-\omega_{b}\\
\sigma^{j}_{\mu\nu} &=& |\mu\rangle_{j}\langle \nu|\\
\Omega_{eg,j}^{w}&=&\frac{1}{2}|e|\langle e|\hat{\textbf{r}}|g\rangle\cdot\hat{\epsilon}^{w}E^{w}(\textbf{r}_{j},t)\label{eq:def1}\\
g_{es,\tau}(\textbf{k})&=&|e|\langle e|\hat{\textbf{r}}|s\rangle\cdot\hat{\epsilon}_{\textbf{k},\tau}f(k)\label{eq:def2}
\end{eqnarray}
We can transform the Hamiltonian into the field interaction picture using the
following unitary transformation:
\begin{eqnarray}
U&=&\exp\Bigg[-\textit{i}\sum_{j=1}^{N_{a}}(\omega^{w}\sigma_{ee}^{j}+\omega_{sg}\sigma_{ss}^{j})t\nonumber\\
&&-\textit{i}\sum_{\tau}\int \frac{d^{3}\textbf{k}}{(2\pi)^{3}}~\omega_{k}a^{\dagger}_{\textbf{k},\tau}a_{\textbf{k},\tau}t\Bigg]
\end{eqnarray}
With this unitary transformation the interaction Hamiltonian is given as:
\begin{equation}
H_{new}=U^{\dagger}H_{old}U+\textit{i}(\partial_{t}U^{\dagger})U
\end{equation}
On solving the expression for $H_{new}$ we get:
\begin{eqnarray}
H_{new}^{w}&=&\sum_{j=1}^{N_{a}}\Delta^{w}\sigma_{ee}^{j}+\sum_{j=1}^{N_{a}}\bigg[\Omega_{eg,j}^{w}e^{\textit{i}\textbf{k}^{w}\cdot\textbf{r}_{j}}\sigma_{eg}^{j}\nonumber\\
&+&\sum_{\tau}\int \frac{d^{3}\textbf{k}}{(2\pi)^{3}}~g_{es,\tau}(\textbf{k})e^{\textit{i}\textbf{k}\cdot\textbf{r}_{j}-\textit{i}(\omega_{k}-\omega^{w}+\omega_{sg})t}\sigma_{es}^{j}a_{\textbf{k},\tau}\nonumber\\
&+&\text{h.c.}\bigg]
\end{eqnarray}
where we have defined $\Delta^{w} = \omega_{eg}-\omega^{w}$ as the detuning of
the write pulse from the $|e\rangle$-$|g\rangle$ transition. We can reduce the three level problem to a two level problem by adiabatic elimination of the excited level $|e\rangle$. This approximation is valid if the natural width
$\Gamma$ of the excited level and frequency spread of the write pulse around
$\omega^{w}$ are significantly smaller compared to the detuning $\Delta^{w}$.\\

The Hamiltonian after the adiabatic elimination thus obtained after ignoring the Stark shifts in level $|s\rangle$ due to spontaneous emission is given by:
\begin{eqnarray}
H_{new}^{w}&=&-\sum_{j=1}^{N_{a}}\frac{|\Omega_{eg,j}^{w}|^{2}}{\Delta^{w}}\sigma_{gg}^{j}-\sum_{j=1,\tau}^{N_{a}}\int \frac{d^{3}\textbf{k}}{(2\pi)^{3}}\nonumber\\
&&\bigg[\frac{\Omega_{eg,j}^{w}g_{es,\tau}^{*}(\textbf{k})}{\Delta^{w}}e^{-\textit{i}(\Delta\textbf{k}\cdot\textbf{r}_{j}-\Delta\omega t)}\sigma_{sg}^{j}a_{\textbf{k},\tau}^{\dagger}\nonumber\\
&&+\text{h.c.}\bigg]
\end{eqnarray}
where:
\begin{eqnarray}
\Delta\textbf{k}&=&\textbf{k}-\textbf{k}^{w}\\
\Delta\omega&=&\omega_{k}-(\omega^{w}-\omega_{sg})
\end{eqnarray}
We can perform another unitary transformation, rotating the vector $|g\rangle$
such that the resulting Hamiltonian depends only on the lowering and raising
atomic operators. The corresponding unitary transformation is:
\begin{equation}
U=\exp\Bigg[\textit{i}\int_{0}^{T_{w}}\sum_{j = 1}^{N_{a}}\frac{|\Omega^{w}_{eg,j}|^{2}}{\Delta^{w}}\sigma_{gg}^{j}dt'\Bigg]
\label{eq:stark}
\end{equation}
The resulting transformed Hamiltonian is then:
\begin{eqnarray}
H_{new}^{w}&=&-\sum_{j=1,\tau}^{N_{a}}\int \frac{d^{3}\textbf{k}}{(2\pi)^{3}}\bigg[\frac{\Omega_{eg,j}^{w}g_{es,\tau}^{*}(\textbf{k})}{\Delta^{w}}e^{-\textit{i}(\Delta\textbf{k}\cdot\textbf{r}_{j}-\Delta\omega t)}\nonumber\\
&&e^{\textit{i}\int_{0}^{T_{w}}\frac{|\Omega^{w}_{eg,j}|^{2}}{\Delta^{w}}dt}\sigma_{sg}^{j}a_{\textbf{k},\tau}^{\dagger}+\text{h.c.}\bigg]
\end{eqnarray}
In the following calculations, we ignore the phase accumulated due to the
Stark shift in $|g\rangle$ as it is small in comparison with the other phases
accumulated in the duration $T_{w}$.\\

Let us start with the write Hamiltonian and derive the state of the system
under the single photon excitation limit. We consider only single photon
excitation as the write laser pulse is weak and off-resonant.
\begin{equation}
H_{new}^{w}=\sum_{j=1,\tau}^{N_{a}} \int \frac{d^{3}\textbf{k}}{(2\pi)^{3}} \big[C_{j,\tau}^{w}(\textbf{k},t) \sigma_{gs}^{j} a_{\textbf{k},\tau} + \text{h.c.}\big] 
\label{eq:writeHamil}
\end{equation}
We have defined:
\begin{eqnarray}
C_{j,\tau}^{w}(\textbf{k},t) & = & - \frac{\Omega_{eg,j}^{w*}g_{es,\tau}(\textbf{k})}{\Delta^{w}}e^{\textit{i}(\Delta\textbf{k}\cdot\textbf{r}_{j}-\Delta\omega t)}
\end{eqnarray}
We consider the write pulse to be a square pulse with a Gaussian transverse profile travelling in the $+z$ direction whose electric field magnitude is given as:
\begin{equation}
E^{w}(\textbf{r},t)=\textit{Q}^{w}(\textbf{r})V^{w}(t)
\label{eq:EF}
\end{equation}

With:
\begin{eqnarray}
\textit{Q}^{w}(\textbf{r}) &=& \frac{E_{0}^{w}}{\sqrt{1+\frac{z^{2}}{z_{w}^2}}}e^{-\frac{x^{2}+y^{2}}{W_{w}^{2}\big(1+\frac{z^{2}}{z_{w}^2}\big)}}e^{\textit{i}\big[\frac{k^{w}(x^{2}+y^{2})}{2R_{w}(z)}-\psi_{w}(z)\big]}
\label{eq:writefield}\\
V^{w}(t) &=& \Theta(t)\Theta(T_{w}-t)
\end{eqnarray}

Where:
\begin{eqnarray}
z_{w}&=&\frac{k^{w}W_{w}^{2}}{2}\\
R_{w}(z)&=&z\left(1+\frac{z_{w}^{2}}{z^{2}}\right)\\
\psi_{w}(z)&=&\tan^{-1}\frac{z}{z_{w}}
\end{eqnarray}

In the above expression, $E_{0}^{w}$ is the peak value of electric field at
the center of the Gaussian profile, $W_{w}$ is the beam waist. According to
the usual convention of defining Gaussian beam we have, $z_{w}$ as the
Rayleigh length, $R_{w}(z)$ as the radius of curvature of the beam wave-front
at the position $z$ and $\psi_{w}(z)$ is the associated Gouy phase.\\

Also in Eq.~(\ref{eq:EF}), we have taken the liberty of expressing the
electric field magnitude as a product of the spatial part and temporal part
since the time taken for the propagation of a single wave-front from one end
of the atomic sample to the other end is very small compared to the total time
duration of the Gaussian square pulse and $T_{w}\omega^{w} \gg1$. For a few
recent experiments where the widths of the control pulses and the single
photon optics is comparable, it becomes necessary to consider the phases
introduced due to the transverse profile of these paraxial pulses
\cite{Pu2017}.\\

A single photon excited state for the write Hamiltonian defined in Eq.~(\ref{eq:writeHamil}) is given as:
\begin{equation}
|\phi\rangle^{w} = \bigg[1 -\textit{i}\int_{0}^{T_{w}} dt ~H^{w}(t)\bigg]|vac\rangle
\end{equation}
where:
\begin{equation}
|vac\rangle = |g\rangle^{\otimes N}|0\rangle_{ph} = |g\rangle_{1}|g\rangle_{2}...|g\rangle_{N}|0\rangle_{ph}
\end{equation}
The state $|0\rangle_{ph}$ stands for the absence of any photons in the system.\\

On substituting the expression for the Hamiltonian, we get:
\begin{eqnarray}
|\phi\rangle^{w} 
&=& |vac\rangle +\frac{|e|^{2}\langle e|\textbf{r}| g\rangle\cdot\hat{\epsilon}^{w}}{16\pi^{3}\Delta^{w}}\frac{1}{\sqrt{2\varepsilon_{0}}}\sum_{j=1,\tau}^{N_{a}} Q^{w}(\textbf{r}_{j})\nonumber\\
&\int d^{3}\textbf{k}&~\langle s|\textbf{r}|e\rangle\cdot\hat{\epsilon}^{*}_{\textbf{k},\tau}\sqrt{\omega_{k}}e^{\textit{i}(k^{w}\hat{\textbf{z}}-\textbf{k})\cdot\textbf{r}_{j}}\nonumber\\
&\int_{0}^{T_{w}}dt&~e^{\textit{i}(\omega_{k}-\omega^{w}+\omega_{sg})t}~\Theta(T_{w}-t)\Theta(t)|s\rangle_{j}a_{\textbf{k},\tau}^{\dagger}|0\rangle_{ph}\nonumber\\
&&\\
&=&|vac\rangle +\frac{|e|^{2}T_{w}\langle e|\textbf{r}| g\rangle\cdot\hat{\epsilon}^{w}}{16\pi^{3}\Delta^{w}}\frac{1}{\sqrt{2\varepsilon_{0}}}\sum_{j=1,\tau}^{N_{a}} Q^{w}(\textbf{r}_{j})\nonumber\\
&\int d^{3}\textbf{k}& \langle s|\textbf{r}|e\rangle\cdot\hat{\epsilon}^{*}_{\textbf{k}\tau}\sqrt{\omega_{k}}e^{\textit{i}(k^{w}\hat{\textbf{z}}-\textbf{k})\cdot\textbf{r}_{j}}e^{\textit{i}(\omega_{k}-\omega^{w}+\omega_{sg})\frac{T_{w}}{2}}\nonumber\\
&&\text{sinc}\bigg[(\omega_{k}-\omega^{w}+\omega_{sg})\frac{T_{w}}{2}\bigg]|s\rangle_{j}a_{\textbf{k},\tau}^{\dagger}|0\rangle_{ph}
\end{eqnarray}
Under the assumption that the single photon detectors used for the detection
of the emitted signal photon are ideal, we can ignore the vacuum component. In
the Schrodinger picture, the above expression can then be understood as:
\begin{eqnarray}
|\phi\rangle^{w}  &=& \sum_{\tau}\int d^{3}\textbf{k}~\hat{f}^{w}(k,\theta_{k},\phi_{k},\tau)e^{-i\omega_{k}t}a_{\textbf{k},\tau}^{\dagger}|0\rangle_{ph}
\end{eqnarray}
where:
\begin{eqnarray}
\hat{f}^{w}(k,\theta_{k},\phi_{k},\tau)&=&\frac{|e|^{2}T_{w}\langle e|\textbf{r}| g\rangle\cdot\hat{\epsilon}^{w}}{16\pi^{3}\Delta^{w}}\sqrt{\frac{\omega_{k}}{2\varepsilon_{0}}}\sum_{j=1}^{N_{a}} Q^{w}(\textbf{r}_{j})\nonumber\\
&& \langle s|\textbf{r}|e\rangle\cdot\hat{\epsilon}^{*}_{\textbf{k}\tau}e^{\textit{i}(k^{w}\hat{\textbf{z}}-\textbf{k})\cdot\textbf{r}_{j}}e^{\textit{i}(\omega_{k}-\omega^{w}+\omega_{sg})\frac{T_{w}}{2}}\nonumber\\
&&\text{sinc}\bigg[(\omega_{k}-\omega^{w}+\omega_{sg})\frac{T_{w}}{2}\bigg]e^{-i\omega_{sg}t}|s\rangle_{j}\nonumber\\
\label{eq:fhatw}
\end{eqnarray}
In the above equation, we do not consider the phase factors coming from
unitary in Eq. (\ref{eq:stark}) as they do not influence the final expression
for IRE . We can now trace over the $\omega_{k}$ component because the single
photon detector is not sensitive to this value. The trace of $|\phi\rangle
^{w}{}^{w}\langle\phi|$ over $\omega_{k}$ diverges for the integration limits
going from 0 to $\infty$, but we can restrict the integration from 0 to a
finite value of frequency based on the validity of the dipole approximation.
For such a situation the dominant contribution comes from a small window
around $\omega_{k}=\omega^{w}-\omega_{sg}$. The remaining angular profile of Eq.~(\ref{eq:fhatw}) becomes:
\begin{eqnarray}
\hat{f}^{w}(\theta_{k},\phi_{k},\tau)&\propto &\sum_{j} Q^{w}(\textbf{r}_{j}) \langle s|\textbf{r}|e\rangle\cdot\hat{\epsilon}^{*}_{\hat{\textbf{k}},\tau}\nonumber\\
&&\times e^{\textit{i}(k^{w}\hat{\textbf{z}}-k^{s}\hat{\textbf{k}})\cdot\textbf{r}_{j}-i\omega_{sg}t}|s\rangle_{j}
\end{eqnarray}
where $\hat{\textbf{k}}$ is the unit wave-vector and 
\begin{eqnarray}
ck^{s} &=& \omega^{w}-\omega_{sg}
\label{eq:signalfreq}
\end{eqnarray}
Experimentally, we couple the emitted photon into a single mode optical fiber which in
turn couples to the single photon detector. The polarization of the emitted
photon is filtered before it is coupled to the optical fiber. The transverse
mode associated with the optical fiber is considered to be a Gaussian mode
propagating in the $+\hat{z}$ direction. The emitted signal photon mode function will be mostly confined in a small angular region around the direction $+\hat{z}$, overlapping with the paraxial
optical fiber mode profile. Thus, we can assume $\hat{\epsilon}^{*}_{\hat{\mathbf{k}%
},\tau} = \hat{\epsilon}^{*}_{\hat{\mathbf{z}},\tau}$ which can now be taken
out of the integration. This approximation is valid since $\hat{\epsilon}%
^{*}_{\hat{\mathbf{k}},\tau}$ varies slowly over the solid angle around
$\hat{\mathbf{z}}$ direction when compared to the rapidly varying phase factor
$e^{-\mathit{i}k^{s}\hat{\mathbf{k}}\cdot\mathbf{r}_{j}}$ with changing
$\hat{\mathbf{k}}$. Also, the polarization, $\tau$, is fixed by the polarization filters. Thus, we have:
\begin{eqnarray}
\hat{f}^{w}(\theta_{k},\phi_{k})&\propto&\sum_{j} Q^{w}(\textbf{r}_{j}) e^{\textit{i}(k^{w}\hat{\textbf{z}}-k^{s}\hat{\textbf{k}})\cdot\textbf{r}_{j}}e^{-i\omega_{sg}t}|s\rangle_{j}\\
&=& N_{f}~\sum_{j=1}^{N_{a}} Q^{w}(\textbf{r}_{j}) e^{\textit{i}(k^{w}\hat{\textbf{z}}-k^{s}\hat{\textbf{k}})\cdot\textbf{r}_{j}}e^{-i\omega_{sg}t}|s\rangle_{j}\nonumber\\
\label{eq:writefunction}
\end{eqnarray}
where $N_{f}$ is the normalization constant for the angular mode function.\newline

The angular mode function of the field associated with the optical fiber can be approximated by a Gaussian mode given
below:
\begin{eqnarray}
 g^{w}(\theta_{k},\phi_{k}) & = &N^{w}_{g}~ e^{-\frac{1}{4}(k^{s}W_{s}\sin\theta_{k})^{2}}
 \label{eq:gauss}
\end{eqnarray}
with $N^{w}_{g}$ as the normalization factor.\\

On taking the overlap between Eq.~(\ref{eq:writefunction}) and Eq.~(\ref{eq:gauss}) in the forward direction we get the spin-wave state $|\phi\rangle^{sw}$ as:
\begin{eqnarray}
 |\phi\rangle^{sw}&=&\int_{0}^{2\pi}d\phi_{k}\int_{0}^{\frac{\pi}{2}}d\theta_{k} \sin\theta_{k} \hat{f}^{w}(\theta_{k},\phi_{k})g^{w*}(\theta_{k},\phi_{k})\nonumber \\
 &&\\
 &=&N^{sw}\sum_{j=1}^{N_{a}} Q^{w}(\textbf{r}_{j})e^{\textit{i}k^{w}z_{j}}\int_{0}^{\frac{\pi}{2}} d\theta_{k}\sin\theta_{k}e^{-\textit{i}k^{s}z_{j}\cos\theta_{k}} \nonumber\\
 &&J_{0}(k^{s}|\textbf{r}_{j\perp}|\sin\theta_{k})e^{-\frac{1}{4}(k^{s}W_{s}\sin\theta_{k})^{2}}e^{-i\omega_{sg}t}|s\rangle_{j}
\end{eqnarray}
where $|\mathbf{r}_{j\perp}| = \sqrt{x_{j}^{2}+y_{j}^{2}}$.\\

For experimental parameters of interest, $k^{s}W_{s}\gg1$. Thus, only a very small
interval of values of $\theta_{k}$ above 0 contributes to the integration,
suggesting that we can make the paraxial approximation. Taking the upper limit
of integration to $\infty$, $\cos\theta_{k} \approx1-\theta_{k}^{2}/2 $ and
$\sin\theta_{k}\approx\theta_{k}$ we get:
\begin{eqnarray}
 |\phi\rangle^{sw} &=&N^{sw}\sum_{j=1}^{N_{a}} Q^{w}(\textbf{r}_{j})e^{\textit{i}(k^{w}-k^{s})z_{j}}\int_{0}^{\infty} d\theta_{k}\theta_{k} \nonumber\\
 &&J_{0}(k^{s}|\textbf{r}_{j\perp}|\theta_{k})e^{-\theta^{2}_{k}[\frac{1}{4}(W_{s}k^{s})^{2}-\frac{\textit{i}}{2}k^{s}z_{j}]}e^{-i\omega_{sg}t}|s\rangle_{j}\nonumber\\
 &&\\
 &=&N^{sw}\sum_{j=1}^{N_{a}} Q^{w}(\textbf{r}_{j})\frac{e^{\textit{i}(k^{w}-k^{s})z_{j}}}{z_{s}k^{s}\sqrt{1+\frac{z_{j}^{2}}{z_{s}^2}}}e^{-\frac{x_{j}^{2}+y_{j}^{2}}{W_{s}^{2}\big(1+\frac{z_{j}^{2}}{z_{s}^2}\big)}}\nonumber\\
 &&e^{-\textit{i}\left[\frac{k^{s}(x_{j}^{2}+y_{j}^{2})}{2R_{s}(z_{j})}-\psi_{s}(z_{j})\right]}e^{-i\omega_{sg}t}|s\rangle_{j}
 \label{eq:inicond}
\end{eqnarray}

where:
\begin{eqnarray}
z_{s}&=&\frac{k^{s}W_{s}^{2}}{2}\\
R_{s}(z_{j})&=&z_{j}\left(1+\frac{z_{s}^{2}}{z_{j}^{2}}\right)\\
\psi_{s}(z_{j})&=&\tan^{-1}\frac{z_{j}}{z_{s}}
\end{eqnarray}
The normalization $N^{sw}$ need not be determined as it corresponds to the
success rate of the write process and does not affect the desired IRE. We now
proceed to the read process, where the spin-wave state is read out and a idler
(read) photon is emitted after a memory storage time interval $T_{m}$.
\subsection{\label{sec:level5}The Read Process}
Let us begin by formulating the read Hamiltonian in a way similar to the write
Hamiltonian. In the read process, a short but strong classical laser pulse on
resonance with the $|e\rangle$-$|s\rangle$ transition is made to interact with
the atomic ensemble. The photon emitted from the $|e\rangle$-$|g\rangle$
transition is collected after polarization filtering. Interaction for the
$|s\rangle$-$|e\rangle$ transition is treated semi-classically and the
spontaneous photon emission from $|e\rangle$-$|g\rangle$ transition is treated quantum mechanically. Assuming dipolar light-matter interactions and the RWA, we can write the read Hamiltonian as:
\begin{eqnarray}
H^{r}&=&\sum_{j=1}^{N_{a}}(\omega_{eg}\sigma_{ee}^{j}+\omega_{sg}\sigma_{ss}^{j})+\sum_{\tau}\int \frac{d^{3}\textbf{k}}{(2\pi)^{3}}\omega_{k}a^{\dagger}_{\textbf{k},\tau}a_{\textbf{k},\tau}\nonumber\\
&+&\sum_{j=1}^{N_{a}}\bigg[\Omega_{es,j}^{r}e^{\textit{i}(\textbf{k}^{r}\cdot\textbf{r}'_{j}-\omega^{r}t)}\sigma_{es}^{j}\nonumber\\
&+&\sum_{\tau}\int \frac{d^{3}\textbf{k}}{(2\pi)^{3}}g_{eg,\tau}(\textbf{k})e^{\textbf{i}\textbf{k}\cdot\textbf{r}'_{j}}\sigma_{eg}^{j}a_{\textbf{k}\tau}+\text{h.c.}\bigg]
\end{eqnarray}
Definitions of $\Omega_{es,j}^{r}$ and $g_{eg,\tau}$ are analogous to the definitions in Eqs. (\ref{eq:def1}-\ref{eq:def2}). The atomic positions may have changed during $T_{m}$, and are denoted by
$\mathbf{r}^{\prime}$.\\

Using the resonance condition for the $|e\rangle$-$|s\rangle$ transition, the
read Hamiltonian in the field interaction picture after the application of the
unitary $U$
\begin{eqnarray}
U&=&\exp\bigg[-\textit{i}\sum_{j=1}^{N_{a}}(\omega^{r}\sigma_{ee}^{j}+\omega_{sg}\sigma_{ss}^{j})t\nonumber\\
&&-\textit{i}\sum_{\tau}\int \frac{d^{3}\textbf{k}}{(2\pi)^{3}}\omega_{k}a^{\dagger}_{\textbf{k},
\tau}a_{\textbf{k},
\tau}t\bigg]
\end{eqnarray}
is given as:
\begin{eqnarray}
H^{r}_{new}&=&\sum_{j=1}^{N_{a}}\omega_{sg}\sigma_{ee}^{j} +\sum_{j=1}^{N_a}\bigg[\Omega_{es,j}^{r}e^{\textit{i}(\textbf{k}^{r}\cdot\textbf{r}'_{j}-\omega_{sg}t)}\sigma_{es}^{j}\nonumber\\
&+&\sum_{\tau}\int\frac{d^{3}\textbf{k}}{(2\pi)^{3}}g_{eg,\tau}(\textbf{k})e^{\textit{i}\textbf{k}\cdot\textbf{r}_{j}-\textit{i}(\omega_{k}-\omega^{r})t}\sigma_{eg}^{j}a_{\textbf{k},\tau}\nonumber\\
&+&\text{h.c.}\bigg]
\end{eqnarray}
We consider the classical read-out pulse to be a square pulse propagating in $-z$ direction with a Gaussian transverse profile and its magnitude given as:
\begin{eqnarray}
E^{r}(\textbf{r})&=&Q^{r}(\textbf{r})V^{r}(t)\label{eq:readef}
\end{eqnarray}
With:
\begin{eqnarray}
Q^{r}(\textbf{r})&=&\frac{E_{0}^{r}}{\sqrt{1+\frac{z^{2}}{z_{r}^{2}}}}e^{-\frac{\textbf{r}_{\perp }^{2}}{W_{r}^{2}\big(1+\frac{z^{2}}{z_{r}^{2}}\big)}}e^{-\textit{i}\big[\frac{k^{r}\textbf{r}_{\perp }^{2}}{2R_{r}(z)}-\psi_{r}(z)\big]}
\label{eq:readfield}\\
z_{r}&=&\frac{k^{r}W_{r}^{2}}{2}\\
R_{r}(z)&=&z\left(1+\frac{z_{r}^{2}}{z^{2}}\right)\\
\psi_{r}(z)&=&\tan^{-1}\frac{z}{z_{r}}\\
V^{r}(t)&=&\Theta(t-T_{p})\Theta(T_{p}+T_{r}-t)
\end{eqnarray}

Here, $T_{p} = T_{m}+T_{w}$ is the duration after which the read pulse is sent measured from the beginning of the write pulse and $T_{r}$ is the duration of the read pulse.\\

Let us consider a general state which satisfies the Schrodinger's equation as follows:
\begin{eqnarray}
|\phi(t)\rangle^{r}&=&\sum_{j=1}^{N_{a}}\big[A_{j}(t)e^{-i\omega_{sg}t}|s\rangle_{j}|0\rangle_{ph}+B_{j}(t)e^{-i\omega^{r}t}|e\rangle_{j}|0\rangle_{ph}\big]\nonumber\\
&&+\sum_{\tau}\int \frac{d^{3}\textbf{k}}{(2\pi)^3} C_{\tau}(\textbf{k},t)e^{-i\omega_{k}t}|g\rangle^{\otimes N}a^{\dagger}_{\textbf{k},\tau}|0\rangle_{ph}
\end{eqnarray}
In the above equation, state $|e\rangle_{j}$ is defined similar to state $|s\rangle_{j}$ is Eq.~(\ref{eq:state-s}). The initial condition for our system is given by Eq.~(\ref{eq:inicond}).\\

Applying the Schrodinger's equation we get:
\begin{eqnarray}
&&\textit{i}\frac{d |\phi(t)\rangle^{r}}{d t}=H^{r}|\phi(t)\rangle^{r}\\
&&\textit{i}\dot{A}_{j}(t) = \Omega_{es,j}^{*r}(t)e^{-\textit{i}(\textbf{k}^{r}\cdot\textbf{r}'_{j}-\omega_{sg}t)}B_{j}(t)\\
&&\textit{i}\dot{B}_{j}(t)=\omega_{sg}B_{j}(t)+\Omega_{es,j}^{r}(t)e^{\textit{i}(\textbf{k}^{r}\cdot\textbf{r}'_{j}-\omega_{sg}t)}A_{j}(t)\nonumber\\
&&~~~~~~~~~+\sum_{\tau}\int \frac{d^{3}\textbf{k}}{(2\pi)^{3}} g_{eg,\tau}(\textbf{k})e^{\textit{i}[\textbf{k}\cdot\textbf{r}'_{j}-(\omega_{k}-\omega^{r})t]}C_{\tau}(\textbf{k},t)\nonumber\\
&&\\
&&\textit{i}\dot{C}_{\tau}(\textbf{k},t)=\sum_{j} g_{eg,\tau}^{*}(\textbf{k})e^{-\textit{i}[\textbf{k}\cdot\textbf{r}'_{j}-(\omega_{k}-\omega^{r})t]}B_{j}(t)
\end{eqnarray}
For simplicity, let us assume the dipole moment associated with the Rabi
frequency $\Omega_{es,j}^{r}(t)$ to be real. This does not change the final
result which only depends on the modulus of this Rabi frequency. Defining:
\begin{eqnarray}
A_{j}(t)&=&e^{\textit{i}\omega_{sg}t}e^{-\textit{i}\textbf{k}^{r}\cdot\textbf{r}'_{j}}e^{\textit{i}\big[\frac{k^{r}\textbf{r}_{\perp j}^{2}}{2R_{r}(z_{j})}-\psi_{r}(z_{j})\big]}\alpha_{j}(t)
\label{eq:alpha}
\end{eqnarray}
Substitute $A_{j}(t)$ as given above in the rate equations.
\begin{eqnarray}
&&\textit{i}\dot{\alpha}_{j}(t) = \omega_{sg}\alpha_{j}(t)+\Omega_{es,j}^{r}(t)B_{j}(t)\\
&&\textit{i}\dot{B}_{j}(t)=\omega_{sg}B_{j}(t)+\Omega_{es,j}^{r}(t)\alpha_{j}(t)\nonumber\\
&&~~~~~~~~+\sum_{\tau}\int \frac{d^{3}\textbf{k}}{(2\pi)^3} g_{eg,\tau}(\textbf{k})e^{\textit{i}[\textbf{k}\cdot\textbf{r}'_{j}-(\omega_{k}-\omega^{r})t]}C_{\tau}(\textbf{k},t)\nonumber\\
&&\label{eq:middle}\\
&&\textit{i}\dot{C}_{\tau}(\textbf{k},t)=\sum_{j} g_{eg,\tau}^{*}(\textbf{k})e^{-\textit{i}[\textbf{k}\cdot\textbf{r}'_{j}-(\omega_{k}-\omega^{r})t]}B_{j}(t)
\label{eq:lasteq}
\end{eqnarray}
Formally integrating Eq.~(\ref{eq:lasteq}) with $C_{\tau}(\textbf{k},T_{p})=0$ we get:
\begin{eqnarray}
C_{\tau}(\textbf{k},t)&=&-\textit{i}\sum_{j}\int_{T_{p}}^{t} dt' g_{eg,\tau}^{*}(\textbf{k})e^{-\textit{i}[\textbf{k}\cdot\textbf{r}'_{j}-(\omega_{k}-\omega^{r})t']}B_{j}(t')\nonumber\\
\label{eq:c-coeff}
\end{eqnarray}
Substituting the above equation into Eq.~(\ref{eq:middle}), we get:
\begin{eqnarray}
\textit{i}\dot{\alpha}_{j}(t) &=& \omega_{sg}\alpha_{j}(t)+\Omega_{es,j}^{r}(t)B_{j}(t)\\
\textit{i}\dot{B}_{j}(t)&=&\omega_{sg}B_{j}(t)+\Omega_{es,j}^{r}(t)\alpha_{j}(t)\nonumber\\
&-& \textit{i}\sum_{l,\tau}\int \frac{d^{3}\textbf{k}}{(2\pi)^{3}} |g_{eg,\tau}(\textbf{k})|^{2}e^{\textit{i}\textbf{k}\cdot(\textbf{r}'_{j}-\textbf{r}'_{l})} \nonumber\\
& & \times\int_{T_{p}}^{t} dt' e^{-\textit{i}(\omega_{k}-\omega^{r})(t-t')}B_{l}(t')\nonumber\\
\end{eqnarray}
Substituting:
\begin{eqnarray}
\tilde{B}_{j}(t)&=&B_{j}(t)e^{\textit{i}\omega_{sg}t}\label{eq:B-co-eff}\\
\tilde{\alpha}_{j}(t)&=&\alpha_{j}(t)e^{\textit{i}\omega_{sg}t}
\end{eqnarray}
We get:
\begin{eqnarray}
\dot{\tilde{\alpha}}&=& -\textit{i}\Omega_{es,j}^{r}(t)\tilde{B}_{j}(t)\\
\dot{\tilde{B}}_{j}(t)&=&-\textit{i}\Omega_{es,j}^{r}(t)\tilde{\alpha}_{j}(t)-\int_{T_{p}}^{t}dt' I_{j}(t,t')\label{eq:integral}
\end{eqnarray}
where: 
\begin{eqnarray}
I_{j}(t,t') = I^{(1)}_{j}(t,t') +I^{(2)}_{j}(t,t')
\label{eq:I}
\end{eqnarray}
with:
\begin{eqnarray}
I^{(1)}_{j}(t,t')&=&\sum_{\tau}\int \frac{d^{3}\textbf{k}}{(2\pi)^3}|g_{eg,\tau}(\textbf{k})|^{2}\nonumber\\
&&e^{-\textit{i}(\omega_{k}-\omega^{r}-\omega_{sg})(t-t')}\tilde{B}_{j}(t')
\label{eq:I1}
\end{eqnarray}
\begin{eqnarray}
I^{(2)}_{j}(t,t')&=&\sum_{\tau}\sum_{l=1,l\neq j}^{N_{a}}\int \frac{d^{3}\textbf{k}}{(2\pi)^3}|g_{eg,\tau}(\textbf{k})|^{2}e^{\textit{i}\textbf{k}\cdot(\textbf{r}'_{j}-\textbf{r}'_{l})}\nonumber\\
&& e^{-\textit{i}(\omega_{k}-\omega^{r}-\omega_{sg})(t-t')}\tilde{B}_{l}(t')\\
&=&\sum_{\tau}\sum_{l=1,l\neq j}^{N_{a}}\int\frac{d^{3}\textbf{k}}{(2\pi)^3}\frac{\omega_{k}}{2\varepsilon_{0}}|\textbf{d}_{eg}\cdot\hat{\epsilon}_{\textbf{k},\tau}|^{2}e^{\textit{i}\textbf{k}\cdot(\textbf{r}'_{j}-\textbf{r}'_{l})}\nonumber\\
&&e^{-\textit{i}(\omega_{k}-\omega^{r}-\omega_{sg})(t-t')}\tilde{B}_{l}(t')\\
&=&\sum_{l=1,l\neq j}^{N_{a}}\int\frac{d^{3}\textbf{k}}{(2\pi)^{3}}\frac{\omega_{k}}{2\varepsilon_{0}}\textbf{d}_{eg}\cdot[I-\hat{\textbf{k}}\hat{\textbf{k}}]\cdot\textbf{d}_{eg}^{*}\nonumber\\
&&e^{\textit{i}\textbf{k}\cdot(\textbf{r}'_{j}-\textbf{r}'_{l})}e^{-\textit{i}(\omega_{k}-\omega^{r}-\omega_{sg})(t-t')}\tilde{B}_{l}(t')\\
&=&\sum_{l=1,l\neq j}^{N_{a}}\int_{0}^{\infty}\frac{dkk^{3}c}{16\pi^{3}\varepsilon_{0}}\int d\Omega_{k}\textbf{d}_{eg}\cdot[I-\hat{\textbf{k}}\hat{\textbf{k}}]\cdot\textbf{d}_{eg}^{*}\nonumber\\
&&e^{\textit{i}\textbf{k}\cdot(\textbf{r}'_{j}-\textbf{r}'_{l})}e^{-\textit{i}(kc-\omega^{r}-\omega_{sg})(t-t')}\tilde{B}_{l}(t')\nonumber\\
&&\\
&=&\sum_{l=1,l\neq j}^{N_{a}}\int_{0}^{\infty}\frac{dkk^{3}c}{4\pi^{2}\varepsilon_{0}}e^{-\textit{i}(kc-\omega^{r}-\omega_{sg})(t-t')}\tilde{B}_{l}(t')\nonumber\\
&&\bigg\{\textbf{d}_{eg}\cdot\bigg[I-\frac{\textbf{r}_{jl}\textbf{r}_{jl}}{|\textbf{r}_{jl}|^{2}}\bigg]\cdot\textbf{d}_{eg}^{*}~j_{0}(k|\textbf{r}_{jl}|)\nonumber\\
&&-\textbf{d}_{eg}\cdot\bigg[I-3\frac{\textbf{r}_{jl}\textbf{r}_{jl}}{|\textbf{r}_{jl}|^{2}}\bigg]\cdot\textbf{d}_{eg}^{*}\frac{j_{1}(k|\textbf{r}_{jl}|)}{k|\textbf{r}_{jl}|}\bigg\}\nonumber
\label{eq:noname}
\\
\end{eqnarray}
Where we have defined:
\begin{eqnarray}
\textbf{r}_{jl} &=& \textbf{r}'_{j}-\textbf{r}'_{l}
\end{eqnarray}
In the above equation, $j_{0}(x)$ and $j_{1}(x)$ are spherical Bessel functions of the first kind. Terms with $j\neq l$ in Eq.~(\ref{eq:noname}) denote atom-atom interactions induced by the quantized electric field which correspond to re-absorption of the emitted photon field. For experimental atomic densities of interest, the average number of atoms separated by a distance of about a $\lambda = 2\pi c/\omega^{r}$ is less than 1. For such low densities we can ignore the re-absorption terms from our calculations, keeping only the terms where $j=l$ in Eq.~(\ref{eq:I}). Then:
\begin{eqnarray}
I_{j}(t,t')&=&I_{j}^{(1)}(t,t')\\
&=&\int_{0}^{\infty}d\omega\frac{\omega^{3}}{6\pi^{2}\varepsilon_{0}c^{3}}|\textbf{d}_{eg}|^{2}e^{-\textit{i}(\omega-\omega^{r}-\omega_{sg})(t-t')}\tilde{B}_{j}(t')\nonumber\\
&&\label{eq:80}\\
&=&\frac{(\omega^{r}+\omega_{sg})^{3}|\textbf{d}_{eg}|^{2}}{6\pi^{2}\varepsilon_{0}c^{3}}2\pi\delta(t-t')\tilde{B}_{j}(t')\\
\label{eq:befgam}
&\equiv& \Gamma_{eg}\delta(t-t')\tilde{B}_{j}(t')
\label{eq:gamma}
\end{eqnarray}
where we use the Wigner-Weisskopf approximation \cite{Berman10}. $\Gamma_{eg}$ is the rate of spontaneous emission from $|e\rangle$ to $|g\rangle$. Substituting Eq.~(\ref{eq:gamma}) into Eq.~(\ref{eq:integral}) we get
\begin{eqnarray}
\dot{\tilde{\alpha}}(t)&=& -\textit{i}\Omega_{es,j}^{r}(t)\tilde{B}_{j}(t)
\label{eq:req1}\\
\dot{\tilde{B}}_{j}(t)&=&-\textit{i}\Omega_{es,j}^{r}(t)\tilde{\alpha}_{j}(t)-\gamma_{eg}\tilde{B}_{j}(t)
\label{eq:req2}
\end{eqnarray}
where $\gamma_{eg} = \Gamma_{eg}/2$.

For the electric field given in Eq.~(\ref{eq:readef}), $\Omega_{es,j}^{r}$ is non-zero only when $T_{p}\leq t\leq T_{p}+T_{r}$. For $ t > T_{p} + T_{r}$:
\begin{eqnarray}
\dot{\tilde{\alpha}}(t)&=& 0\\
\dot{\tilde{B}}_{j}(t)&=&-\gamma_{eg}\tilde{B}_{j}(t)
\end{eqnarray}
Thus, for $t > T_{p}+T_{r}$
\begin{eqnarray}
\tilde{\alpha}(t)&=& \tilde{\alpha}(T_{p}+T_{r})\label{eq:alphaprim}\\
\tilde{B}_{j}(t)&=&\tilde{B}_{j}(T_{p}+T_{r})~e^{-\gamma_{eg}(t-T_{p}-T_{r})}
\label{eq:betaprim}
\end{eqnarray}
Now let us evaluate the solution to the rate equations (Eqs. \ref{eq:req1}-\ref{eq:req2}) for $T_{p}\leq t \leq T_{p}+T_{r}$. This set of two first order differential equations can be combined into a single second order differential equation given as:
\begin{eqnarray}
\ddot{\tilde{B}}_{j}(t)&=&-(\Omega_{es,j}^{r})^{2}\tilde{B}_{j}(t)-\gamma_{eg}\dot{\tilde{B}}_{j}(t)
\label{eq:dif}
\end{eqnarray}

Let $\tilde{\Omega}_{es,j}\equiv \sqrt{(\Omega^{r}_{es,j})^{2}-\frac{\gamma_{eg}^{2}}{4}}$. The solution to the Eq.~(\ref{eq:dif}) is:
\begin{eqnarray}
\tilde{B}_{j}(t)=C_{1}e^{(-\frac{\gamma_{eg}}{2}-\textit{i}\tilde{\Omega}_{es,j})t}+C_{2}e^{(-\frac{\gamma_{eg}}{2}+\textit{i}\tilde{\Omega}_{es,j})t}
\end{eqnarray}
Using the initial conditions at $t = T_{p}$ we get:
\begin{eqnarray}
C_{1}&=&\frac{\Omega_{es,j}^{r}}{2\tilde{\Omega}_{es,j}}\alpha_{j}(T_{P})e^{(\frac{\gamma_{eg}}{2}+\textit{i}\tilde{\Omega}_{es,j}+\textit{i}\omega_{sg})T_{p}}\label{eq:c1-coeff}\\
C_{2}&=&-\frac{\Omega_{es,j}^{r}}{2\tilde{\Omega}_{es,j}}\alpha_{j}(T_{P})e^{(\frac{\gamma_{eg}}{2}-\textit{i}\tilde{\Omega}_{es,j}+\textit{i}\omega_{sg})T_{p}}\label{eq:c2-coeff}
\end{eqnarray}
Evaluating $C_{\textbf{k},\tau}(t)$ (Eq.~\ref{eq:c-coeff}) using Eq.~(\ref{eq:B-co-eff}) and Eqs.~(\ref{eq:c1-coeff}-\ref{eq:c2-coeff}) with the definition $\Delta_{k}^{r}\equiv(\omega_{k}-\omega^{r}-\omega_{sg})$ we get:
\begin{widetext}
\begin{eqnarray}
C_{\tau}(\textbf{k},t)&=&-\sum_{j} g_{eg,\tau}^{*}(\textbf{k})\frac{\Omega_{es,j}^{r}}{\tilde{\Omega}_{es,j}}\alpha_{j}(T_{p})e^{(\frac{\gamma_{eg}}{2}+\textit{i}\omega_{sg})T_{p}}e^{-\textit{i}\textbf{k}\cdot\textbf{r}'_{j}}\int_{T_{p}}^{t} dt'e^{\textit{i}(\omega_{k}-\omega^{r}-\omega_{sg})t'-\frac{\gamma_{eg}}{2}t'}\sin[\tilde{\Omega}_{es,j}(t'-T_{p})]\nonumber\\
&=&-\sum_{j} g_{eg,\tau}^{*}(\textbf{k})\Omega_{es,j}^{r}\alpha_{j}(T_{p})e^{\textit{i}(\omega_{k}-\omega^{r})T_{p}}e^{-\textit{i}\textbf{k}\cdot\textbf{r}'_{j}}\nonumber\\
&&\times\frac{e^{(-\frac{\gamma_{eg}}{2}+\textit{i}\Delta_{k}^{r})(t-T_{p})}\left\{\cos[\tilde{\Omega}_{es,j}(t-T_{p})]-\frac{\textit{i}\Delta_{k}^{r}-\gamma_{eg}/2}{\tilde{\Omega}_{es,j}}\sin[\tilde{\Omega}_{es,j}(t-T_{p})]\right\}-1}{(\Delta_{k}^{r}-\tilde{\Omega}_{es,j}+\textit{i}\frac{\gamma_{eg}}{2})(\Delta_{k}^{r}+\tilde{\Omega}_{es,j}+\textit{i}\frac{\gamma_{eg}}{2})}\nonumber\\
\end{eqnarray}
\end{widetext}
At $t=T_{p}+T_{r}$ we get:
\begin{widetext}
\begin{eqnarray}
C_{\tau}(\textbf{k},T_{p}+T_{r})&=&-\sum_{j} g_{eg,\tau}^{*}(\textbf{k})\Omega_{es,j}^{r}\alpha_{j}(T_{p})e^{\textit{i}(\omega_{k}-\omega^{r})T_{p}}e^{-\textit{i}\textbf{k}\cdot\textbf{r}'_{j}}\frac{e^{(-\frac{\gamma_{eg}}{2}+\textit{i}\Delta_{k}^{r})T_{r}}\left[\cos(\tilde{\Omega}_{es,j}T_{r})-\frac{\textit{i}\Delta_{k}^{r}-\gamma_{eg}/2}{\tilde{\Omega}_{es,j}}\sin(\tilde{\Omega}_{es,j}T_{r})\right]-1}{(\Delta_{k}^{r}-\tilde{\Omega}_{es,j}+\textit{i}\frac{\gamma_{eg}}{2})(\Delta_{k}^{r}+\tilde{\Omega}_{es,j}+\textit{i}\frac{\gamma_{eg}}{2})}\nonumber\\
\end{eqnarray}
\end{widetext}

We can now find the explicit expression for $C_{\tau}(\textbf{k},t)$ when $t > T_{p}+T_{r}$:
\begin{eqnarray}
C_{\tau}(\textbf{k},t)&=&C_{\tau}(\textbf{k},T_{p}+T_{r})-\textit{i}\sum_{j}\int_{T_{p}+T_{r}}^{t} dt'\nonumber\\
&& g_{eg,\tau}^{*}(\textbf{k})e^{-\textit{i}[\textbf{k}\cdot\textbf{r}'_{j}-(\omega_{k}-\omega^{r})t']}B_{j}(t')
\end{eqnarray}
\begin{eqnarray}
C_{\tau}(\textbf{k},t)&=&C_{\tau}(\textbf{k},T_{p}+T_{r})-\sum_{j}g_{eg,\tau}^{*}(\textbf{k})\frac{\Omega_{es,j}^{r}}{\tilde{\Omega}_{es,j}}\alpha_{j}(T_{p})\nonumber\\
&&e^{-(\frac{\gamma_{eg}}{2}+\textit{i}\omega_{sg})T_{r}}\sin(\tilde{\Omega}_{es,j}T_{r})\nonumber\\
&&e^{-\textit{i}\textbf{k}\cdot\textbf{r}'_{j}}e^{\textit{i}\omega_{sg}(T_{p}+T_{r})}e^{\gamma_{eg}(T_{p}+T_{r})}\nonumber\\
&&\times \frac{e^{(\textit{i}\Delta_{k}^{r}-\gamma_{eg})t}-e^{(\textit{i}\Delta_{k}^{r}-\gamma_{eg})(T_{p}+T_{r})}}{\textit{i}\Delta_{k}^{r}-\gamma_{eg}}\nonumber\\
\end{eqnarray}
Finally:
\begin{eqnarray}
C_{\tau}(\textbf{k},t)&=&-\sum_{j} g_{eg,\tau}^{*}(\textbf{k})\Omega_{es,j}^{r}\alpha_{j}(T_{p})e^{-\textit{i}\textbf{k}\cdot\textbf{r}'_{j}}\nonumber\\
&\Bigg\{&e^{\textit{i}\Delta_{k}^{r}(T_{p}+T_{r})}
e^{\textit{i}\omega_{sg}T_{p}}e^{-\frac{\gamma_{eg}}{2}T_{r}}\nonumber\\
&&\Bigg[ \frac{\cos(\tilde{\Omega}_{es,j}T_{r})-\frac{\textit{i}\Delta_{k}^{r}-\gamma_{eg}/2}{\tilde{\Omega}_{es,j}}\sin(\tilde{\Omega}_{es,j}T_{r})}{(\Delta_{k}^{r}-\tilde{\Omega}_{es,j}+\textit{i}\frac{\gamma_{eg}}{2})(\Delta_{k}^{r}+\tilde{\Omega}_{es,j}+\textit{i}\frac{\gamma_{eg}}{2})}\nonumber\\
&&+\frac{\sin(\tilde{\Omega}_{es,j}T_{r})}{\tilde{\Omega}_{es,j}}\frac{e^{(\textit{i}\Delta_{k}^{r}-\gamma_{eg})(t-T_{p}-T_{r})}-1}{\textit{i}\Delta_{k}^{r}-\gamma_{eg}}\Bigg]\nonumber\\
&-&\frac{e^{\textit{i}(\Delta_{k}^{r}+\omega_{sg})T_{p}}}{(\Delta_{k}^{r}+\textit{i}\frac{\gamma_{eg}}{2}-\tilde{\Omega}_{es,j})(\Delta_{k}^{r}+\textit{i}\frac{\gamma_{eg}}{2}+\tilde{\Omega}_{es,j})}\Bigg\}\nonumber\\
\end{eqnarray}
Substituting $\alpha_{j}(T_{p})$ back using Eq.~(\ref{eq:alpha}) and defining: 
\begin{eqnarray}
\phi^{r}(\textbf{r}'_{j})&=&\frac{k^{r}\textbf{r}_{\perp j}^{'2}}{2R_{r}(z'_{j})}-\psi_{r}(z'_{j})
\end{eqnarray}
we get:
\begin{eqnarray}
C_{\tau}(\textbf{k},t)&=&-\sum_{j} g_{eg,\tau}^{*}(\textbf{k})\Omega_{es,j}^{r}A_{j}(T_{p})e^{\textit{i}\textbf{k}^{r}\cdot\textbf{r}'_{j}}e^{-\textit{i}\phi^{r}(\textbf{r}'_j)}\nonumber\\
&&\times e^{-\textit{i}\textbf{k}\cdot\textbf{r}'_{j}}\zeta(\omega_{k},\textbf{r}'_{j},t)
\label{eq:finl}
\end{eqnarray}
where:
\begin{eqnarray}
\zeta(\omega_{k},\textbf{r}'_{j},t)&=&e^{\textit{i}\Delta_{k}^{r}(T_{p}+T_{r})}e^{-\frac{\gamma_{eg}}{2}T_{r}}\nonumber\\
&\Bigg[& \frac{\cos(\tilde{\Omega}_{es,j}T_{r})-\frac{\textit{i}\Delta_{k}^{r}-\frac{\gamma_{eg}}{2}}{\tilde{\Omega}_{es,j}}\sin(\tilde{\Omega}_{es,j}T_{r})}{(\Delta_{k}^{r})^2-(\Omega^{r}_{es,j})^2+\textit{i}\gamma_{eg}\Delta_{k}^{r}}\nonumber\\
&&+\frac{\sin(\tilde{\Omega}_{es,j}T_{r})}{\tilde{\Omega}_{es,j}}\frac{e^{(\textit{i}\Delta_{k}^{r}-\gamma_{eg})(t-T_{p}-T_{r})}-1}{\textit{i}\Delta_{k}^{r}-\gamma_{eg}}\Bigg]\nonumber\\
&-&\frac{e^{\textit{i}\Delta_{k}^{r}T_{p}}}{(\Delta_{k}^{r})^2-(\Omega^{r}_{es,j})^2+\textit{i}\gamma_{eg}\Delta_{k}^{r}}
\label{eq:zeta}
\end{eqnarray}
At this point another simplification can be made by taking the experimental
conditions into consideration. The read-out pulse generally has a very broad
waist size compared to the write pulse i.e. $W_{r} \gg W_{w}$. In this case,
we can assume that the Gaussian read-out pulse is spatially broad enough to
neglect the dependence of $\Omega_{es,j}^{r}$ on atomic positions. Similarly, we can neglect the phase contributions $\phi^{r}(\mathbf{r}^{\prime}_{j})$.
Also, we assume that $\Omega_{es}^{r} > \frac{\gamma_{eg}}{2}$. \newline

The last term of Eq.~(\ref{eq:zeta}) is the only term that doesn't have the
decay contributions from the excited level. From the experimental perspective,
we can choose $\gamma_{eg}T_{r} \gg1 $, thus we can neglect the first two
terms:
\begin{eqnarray}
\zeta(\omega_{k},\textbf{r}'_{j},t)&\approx&-\frac{e^{\textit{i}\Delta_{k}^{r}T_{p}}}{(\Delta_{k}^{r})^2-(\Omega^{r}_{es})^2+\textit{i}\gamma_{eg}\Delta_{k}^{r}}
\end{eqnarray}
Incorporating these approximations we have:
\begin{eqnarray}
C_{\tau}(\textbf{k},t)&=&\Omega_{es}^{r}\sum_{j}g_{eg,\tau}^{*}(\textbf{k})A_{j}(T_{p})e^{\textit{i}\textbf{k}^{r}\cdot\textbf{r}'_{j}}e^{-\textit{i}\textbf{k}\cdot\textbf{r}'_{j}}\nonumber\\
&&\times \frac{e^{\textit{i}\Delta_{k}^{r}T_{p}}}{(\Delta_{k}^{r})^2-(\Omega^{r}_{es})^2+\textit{i}\gamma_{eg}\Delta_{k}^{r}}
\end{eqnarray}
After sufficiently long time interval only the $C_{\tau}(\textbf{k},t)$ co-efficient survives. Thus, the final state after the action of the read Hamiltonian can be written as:
\begin{eqnarray}
|\Phi\rangle^{r}&=&\sum_{\tau}\frac{1}{8\pi^{3}} \int d^{3}\textbf{k} C_{
\tau}(\textbf{k},t)e^{-i\omega_{k}t}|g\rangle^{\otimes N}a_{\textbf{k}\tau}^{\dagger}|0\rangle_{ph}\nonumber\\
\label{eq:readmode}
\end{eqnarray}
We see that the mode function in Eq.~(\ref{eq:readmode}) peaks for a small
range of values of $\omega_{k}$. We can take the frequency at which the photon
gets emitted by setting $\Delta_{k}^{r}\pm\Omega^{r}_{es} = 0$. Since $\omega^{r}\gg\Omega^{r}_{es}$, taking $\Delta_{k}^{r} = 0$ is a good
approximation. Then by tracing over the frequency part we can now write the
angular part of the emitted photon as:
\begin{eqnarray}
\hat{f}^{r}(\theta_{k},\phi_{k},\tau)&=&\sum_{j}g_{eg,\tau}^{*}(\theta_{k},\phi_{k})A_{j}(T_{p})e^{\textit{i}\textbf{k}^{r}\cdot\textbf{r}'_{j}}e^{-\textit{i}k^{i}\hat{\textbf{k}}\cdot\textbf{r}'_{j}}\nonumber\\
&&\frac{\omega^{r}+\omega_{sg}}{\sqrt{8\pi^{2}\gamma_{eg}c^{3}}}|g\rangle^{\otimes N}
\label{eq:tranread}
\end{eqnarray}
where:
\begin{eqnarray}
ck^{i}&=& \omega^{r}+\omega_{sg}
\label{eq:idlerfreq}
\end{eqnarray}
Using arguments similar to those used in the write part we assume
$g^{*}_{ge,\tau}(\theta_{k},\phi_{k})$ varies slowly for the
relevant values of $\theta_{k},\phi_{k}$ around $\theta_{k} = \pi$. Thus, fixing the wave-vector
direction to be $-\hat{\mathbf{z}}$, as was done for the write process, we can
find the overlap between the angular profile of the emitted photon and the
optical fiber used to collect it. The polarization also gets fixed by the
polarization filter before coupling into the optical fiber. We can also ignore
the phase factors associated with time evolution as the final IRE expression
is independent of it. Note that Eq.~(\ref{eq:tranread}) has the same
normalization as $A_{j}(T_{p})$:
\begin{eqnarray}
\int d\Omega_{k} |\hat{f}^{r}(\theta_{k},\phi_{k})|^{2} &=& \sum_j |A_j(T_p)|^2
\end{eqnarray}
Here we calculate the normalization factor only for the completeness of the
formula. In the numerical simulation it is much easier to directly sample the
angular dependence and then normalize the function, because $g_{eg,\tau}$ is
taken as constant. See Sec.~\ref{sec:level6} for more details. Let the angular profile of the electric field associated with the optical fiber be given as:
\begin{equation}
g^{r}(\theta_{k},\phi_{k})=N_{g}^{r}e^{-\frac{1}{4}(k^{i} W_{i} \sin
\theta_{k})^{2}}\label{eq:idler}
\end{equation}
In the calculation of the overlap we again 
use the paraxial approximation due
to the fact that $k^{i} W_{i}\gg1$. The normalization factor $N_{g}^{R}$ under
this approximation is given as $N_{g}^{r}= k^{i}W_{i}/\sqrt{2}$. Taking the overlap of the emitted photon profile with the Gaussian collection mode then gives the final atomic state:
\begin{eqnarray}
|\phi\rangle^{fs} &=& \int_{0}^{2\pi}d\phi_{k}\int_{\frac{\pi}{2}}^{\pi}d\theta_{k} \sin\theta_{k} \hat{f}^{r}(\theta_{k},\phi_{k})g^{r*}(\theta_{k},\phi_{k})\nonumber \\
&&\\
&=&\frac{\omega^{r}+\omega_{sg}}{\sqrt{8\pi^{2}\gamma_{eg}c^{3}}}\frac{W_{i}k^{i}}{\sqrt{2}}\sum_{j=1}^{N_{a}} A_{j}(T_{p})e^{-\textit{i}k^{r}z'_{j}}\int_{0}^{\frac{\pi}{2}} d\theta_{k}\sin\theta_{k} \nonumber\\
&&e^{\textit{i}k^{i}z'_{j}\cos\theta_{k}}J_{0}(k^{i}|\textbf{r}'_{j\perp}|\sin\theta_{k})e^{-\frac{1}{4}(k^{i}W_{i}\sin\theta_{k})^{2}}|g\rangle^{\otimes N}\nonumber\\
&&\\
&=&\frac{(\omega^{r}+\omega_{sg})}{\sqrt{8\pi^{2}\gamma_{eg}c^{3}}}\frac{W_{i}k^{i}}{\sqrt{2}}\sum_{j=1}^{N_{a}} A_{j}(T_{p})e^{-\textit{i}k^{r}z'_{j}}\frac{e^{\textit{i}k^{i}z'_{j}}}{\sqrt{1+\frac{z_{j}^{'2}}{z_{i}^2}}}\nonumber\\
&&e^{-\frac{x_{j}^{'2}+y_{j}^{'2}}{W_{i}^{2}\big(1+\frac{z_{j}^{'2}}{z_{i}^2}\big)}}e^{\textit{i}\big[\frac{k^{i}(x_{j}^{'2}+y_{j}^{'2})}{2R_{i}(z'_{j})}-\psi_{i}(z'_{j})\big]}|g\rangle^{\otimes N}\nonumber\\
\end{eqnarray}
\begin{eqnarray}
|\phi\rangle^{fs}&\equiv&\sum_j \Lambda(\textbf{r}_{j},\textbf{r}'_{j})|g\rangle^{\otimes N}
\label{eq:finalstate}
\end{eqnarray}
where:
\begin{eqnarray}
z_{i}&=&\frac{k^{i}W_{i}^{2}}{2}\\
R_{i}(z'_{j})&=&z'_{j}\left(1+\frac{z_{i}^{2}}{z_{j}^{'2}}\right)\\
\psi_{i}(z'_{j})&=&\tan^{-1}\frac{z'_{j}}{z_{i}}
\end{eqnarray}
Any subscript or superscript `i' stands for the idler photon. The IRE, $\eta$, is given by the modulus squared of the above overlap. 
\begin{eqnarray}
\eta&=& \frac{\big|\sum_{j} \Lambda(\textbf{r}_{j},\textbf{r}'_{j})\big|^{2}}{\sum_j |A_j(T_p)|^2}
\end{eqnarray}
For an explicit expression for $\Lambda(\textbf{r}_{j},\textbf{r}'_{j})$, we substitute $A_{j}(T_{p})$ from Eq.~(\ref{eq:inicond}), with its normalization factors neglected:
\begin{widetext}
\begin{eqnarray}
\Lambda(\textbf{r}_{j},\textbf{r}'_{j})&=& \frac{\omega^{r}+\omega_{sg}}{\sqrt{8\pi^{2}\gamma_{eg}c^{3}}}\frac{W_{i}k^{i}}{\sqrt{2}}e^{-\textit{i}k^{r}z'_{j}}\frac{e^{\textit{i}k^{i}z'_{j}}}{\sqrt{1+\frac{z_{j}^{'2}}{z_{i}^2}}}\frac{e^{-\textit{i}k^{s}z_{j}}}{\sqrt{1+\frac{z_{j}^{2}}{z_{s}^2}}}\frac{e^{\textit{i}k^{w}z_{j}}}{\sqrt{1+\frac{z_{j}^{2}}{z_{w}^2}}}e^{-\frac{x_{j}^{2}+y_{j}^{2}}{W_{w}^{2}\big(1+\frac{z_{j}^{2}}{z_{w}^2}\big)}}e^{-\frac{x_{j}^{2}+y_{j}^{2}}{W_{s}^{2}\big(1+\frac{z_{j}^{2}}{z_{s}^2}\big)}}e^{-\frac{x_{j}^{'2}+y_{j}^{'2}}{W_{I}^{2}\big(1+\frac{z_{j}^{'2}}{z_{i}^2}\big)}}\nonumber\\
 &&e^{\textit{i}\big[\frac{k^{w}(x_{j}^{2}+y_{j}^{2})}{2R_{w}(z_{j})}-\psi_{w}(z_{j})\big]}e^{-\textit{i}\big[\frac{k^{s}(x_{j}^{2}+y_{j}^{2})}{2R_{s}(z_{j})}-\psi_{s}(z_{j})\big]}e^{\textit{i}\big[\frac{k^{i}(x_{j}^{'2}+y_{j}^{'2})}{2R_{i}(z'_{j})}-\psi_{i}(z'_{j})\big]}
 \label{eq:statefs}
\end{eqnarray}
\end{widetext}
As seen from Eq.~(\ref{eq:statefs}), the coefficient of the the ground state
is a result of weighted interference effects between all the atoms in the
ensemble. The overall effect is equivalent to the overlap of four Gaussian
beams with different beam parameters. Incidentally, the phase-matching
condition cannot be perfectly satisfied even if atoms are stationary as well as for
colinear beams. Substituting the values of $k_{s}$ and $k_{i}$ from
Eq.~(\ref{eq:signalfreq}) and Eq.~(\ref{eq:idlerfreq}) respectively into
Eq.~(\ref{eq:statefs}), we see that there is always a non-zero phase
contribution along the $z$ axis due to $\omega_{sg}$. More precisely, the
coherent atomic spin wave has a wavelength of about $2\pi c/(2\omega_{sg})$ in
the $z$ direction. For ${}^{87}\mathrm{Rb}$ the hyperfine splitting
$|\omega_{sg}| = 2\pi\times6.8\,$GHz, which means $2\pi c/(2\omega_{sg})
\approx$ 22mm. Nevertheless, most experiments never use atomic samples having
sizes lager than a few mm, so this effect will be small. The Gaussian
transverse structure is another contributor that prevents the IRE from being
unity.\\

Let us now use this framework to look at IRE calculated from a numerical simulation of an atomic sample that mimics the write-read process for realistic experimental setup to gain further insight.

\section{\label{sec:level6}Numerical Analysis}
\begin{figure}[h]
  \includegraphics[width=\linewidth]{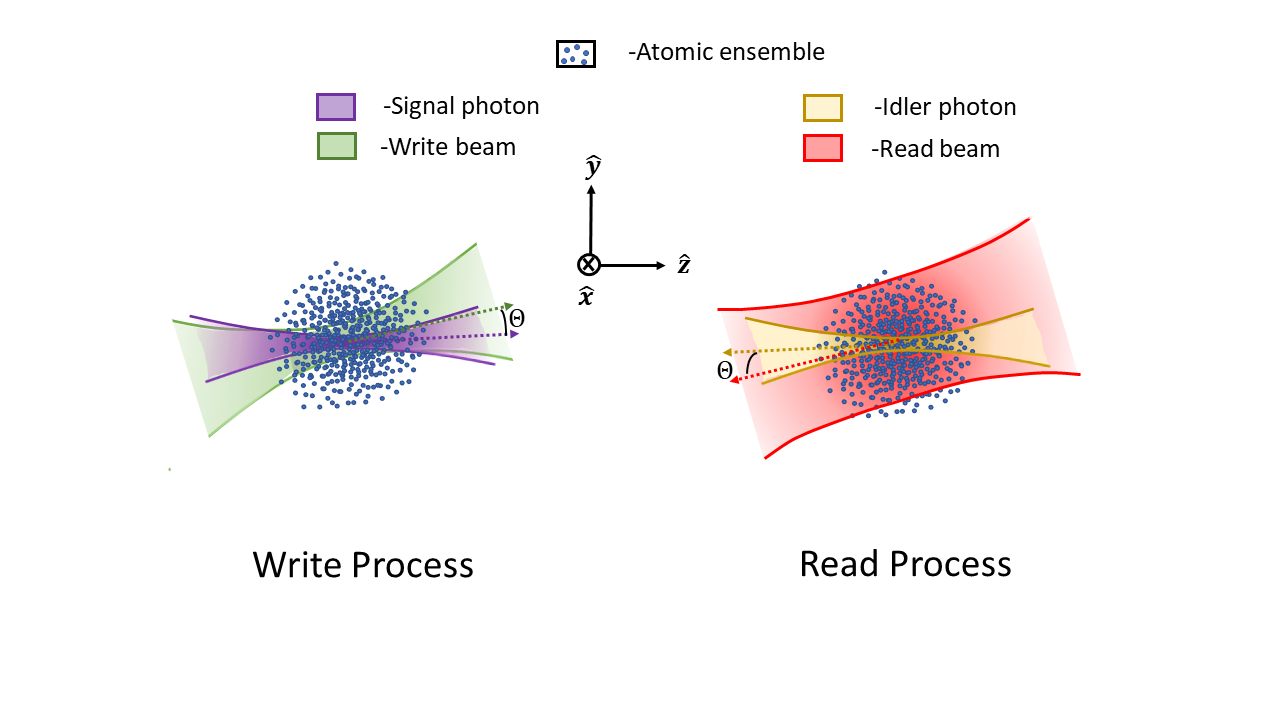}
  \caption{(color online). Configuration of write-read process. Write process: The atomic ensemble is first excited with a classical write pulse, and the emitted signal photon is collected by an optical fiber rotated by an angle $\Theta$ with respect to write pulse axis. The centers of the atomic ensemble and both the beams are aligned. The write beam is generally broader than the signal photon collection beam. Read process: After the write process the ensemble is excited with a very broad read beam which is rotated by an angle $\Theta$ with respect to the idler photon collection beam.  }
  \label{fig:atmcon}
\end{figure}

To avoid the noise associated with detection of the classical write and read
pulses instead of emitted signal and idler photons, a skewed beam
configuration of the write and read beams is implemented experimentally as is
shown in Fig.~(\ref{fig:atmcon}) (\cite{Pu2017},\cite{Sangouard11}%
,\cite{Zhao09},\cite{Yang16},\cite{Simon07}). The write and read laser pulses
aligned along the same axis are rotated by a small angle $\Theta$ with respect
to the alignment axis of the signal and idler collection ports. This can be
easily incorporated into our expression of $\eta$. Assume that the expressions
for the write and read pulse electric field in Eq.~(\ref{eq:writefield}) and
Eq.~(\ref{eq:readfield}) is evaluated in a frame of reference rotated along
the x-axis by a skew angle $\Theta$ such that the beams propagate along the
$\tilde{z}$-direction of this new frame. The signal and idler photon beams
propagate along the z-axis in the original frame of reference. We can express
the write and read beams in the un-rotated frame of reference by making the
following transformations:
\begin{align}
\tilde{x} & =x\\
\tilde{y} & =y\cos\Theta-z\sin\Theta\\
\tilde{z} & =y\sin\Theta+z\cos\Theta
\end{align}
Here the coordinates with tilde denote those in the rotated frame expressed in
terms of the coordinates in the original frame of reference. With this given
transformation, we get:
\begin{widetext}
\begin{eqnarray}
\Lambda(\textbf{r}_{j},\textbf{r}'_{j})&=&  \frac{(\omega^{r}+\omega_{sg})}{\sqrt{8\pi^{2}\gamma_{eg}c^{3}}}\frac{W_{i}k^{i}}{\sqrt{2}}\frac{e^{\textit{i}k^{i}z'_{j}} e^{-\textit{i}k^{s}z_{j}} e^{\textit{i}k^{w}(y_{j}\sin\Theta+z_{j}\cos\Theta)}e^{-\textit{i}k^{r}(y'_{j}\sin\Theta+z'_{j}\cos\Theta)}}{\sqrt{1+\frac{z_{j}^{'2}}{z_{i}^2}} \sqrt{1+\frac{z_{j}^{2}}{z_{s}^2}}\sqrt{1+\frac{(y_{j}\sin\Theta+z_{j}\cos\Theta)^{2}}{z_{w}^2}}}e^{\textit{i}(\psi_{s}(z_{j})-\psi_{i}(z'_{j})-\psi_{w}(y_{j}\sin\Theta+z_{j}\cos\Theta)]}\nonumber\\
 &&e^{-\frac{x_{j}^{2}+y_{j}^{2}}{W_{w}^{2}\big(1+\frac{z_{j}^{2}}{z_{w}^2}\big)}}e^{-\frac{x_{j}^{2}+y_{j}^{2}}{W_{s}^{2}\big(1+\frac{z_{j}^{2}}{z_{s}^2}\big)}}e^{-\frac{x_{j}^{'2}+(y'_{j}\cos\Theta-z'_{j}\sin\Theta)^{2}}{W_{i}^{2}\big[1+\frac{(y'_{j}\sin\Theta+z'_{j}\cos\Theta)^{2}}{z_{i}^2}\big]}}e^{\textit{i}\big\{\frac{k^{w}[x_{j}^{2}+(y_{j}\sin\Theta+z_{j}\cos\Theta)^{2}]}{2R_{w}(y_{j}\sin\Theta+z_{j}\cos\Theta)}-\frac{k^{s}(x_{j}^{2}+y_{j}^{2})}{2R_{s}(z_{j})}+\frac{k^{i}(x_{j}^{'2}+y_{j}^{'2})}{2R_{i}(z'_{j})}\big\}}
 \label{eq:bigeq}
\end{eqnarray}
\end{widetext}
Throughout the numerical analysis we will assume a Gaussian distribution of
atoms inside a MOT. After the atoms have been cooled by using cyclic cooling
and optical gradient cooling, the atomic sample has a standard deviation of
$0.75\,$mm and the temperature of the atomic sample is about tens of $\mu$K.
We get a most probable speed $\sqrt{\frac{2k_{B}T}{M}}$ which is about a
few cm/s. For Rb atoms with mass $M = 87\, a.u.$ at the temperature of
$30\,\mu K$, this value is about $7.5\,$cm/s. For the time duration when the
spin wave is stored in the atomic ensemble, atomic motion causes degradation
of coherence. We introduce this effect in our calculations by assuming
ballistic motion of atoms:
\begin{eqnarray}
\textbf{r}'_{j}&=&\textbf{r}_{j}+\textbf{v}_{j}T_{m}
\label{eq:newpos}
\end{eqnarray}
where $\mathbf{v}_{j}$ are drawn from a Maxwell-Boltzmann distribution of
velocities. Since the atomic density is not very high, we can ignore
collisions. We have neglected the motion of atoms when the write and read
pulses interact with the atomic ensemble, since they are short enough to
assume that the atoms are stationary for $T_{p}$ and $T_{r}$. The expression
for $\eta$ with the velocities included can be derived by substituting
Eq.~(\ref{eq:newpos}) into Eq.~(\ref{eq:bigeq}).\\

From this equation it becomes clear that the decoherence effect
for a non-zero storage time is a direct result of the atomic motion. Let us
look at the behaviour of the IRE as a function of the different experimental
parameters obtained from a Monte-Carlo sampling of a Gaussian atomic ensemble
with spherical symmetry. The range of parameters chosen for all the numerical
simulation henceforth have been inspired by experiments reported in Ref.~\cite{Pu2017}. The atomic samples generated for the numerical simulations have a peak density of the order of $10^{17}$ atoms/$\text{m}^{3} $. An important quantity that captures the strength of interaction between the atomic ensemble and the light is the optical depth of the ensemble. For a given Gaussian density profile the optical depth for a sample of atoms interacting with Gaussian beams is given by the following expression:
\begin{eqnarray}
OD &=& \frac{2}{\pi }\int^{\infty}_{-\infty} dz \frac{2\pi c_{CG}^{2}\sigma_{0}}{w_{0}^2(1+\frac{z^{2}}{z_{w}^{2}})}\int_{0}^{\infty} rdr n_{0}e^{-\frac{r^{2}+z^{2}}{2r_{0}^{2}}}e^{-\frac{2r^2}{w_{0}^{2}(1+\frac{z^2}{z_{w}^2})}}\nonumber\\
\end{eqnarray}
where $w_{0}$ is the Gaussian beam waist at $z=0$, $\sigma_{0} $ the atomic
cross-section, $n_{0}$ the peak atomic density and $r_{0}$ as the standard
deviation of the atomic distribution. $z_{w}$ is the Rayleigh length for the
Gaussian beam given as $k_{0} w_{0}^{2}/2$ for wave-number $k_{0}$.
$c_{CG}^{2}$ is the square of the Clebsch-Gordon coefficient associated with
the particular atomic transition of interest. We will calculate the optical
depth for the interaction with an off-resonant write-pulse corresponding to
the 795nm D1 line in ${}^{87}\text{Rb}$. The off-resonant cross-section for
this transition is $\sigma_{0} = 1.082*10^{-9}\text{cm}^{-2}$ \cite{Steck09}. For convenience, we set $c_{CG} = 1$. The OD can be scaled with the
appropriate value of $c_{CG}$ if necessary. For all the numerical results presented in this section, we use $\Delta= 2\pi\times\text{10}$MHz and $\omega_{sg} = -2\pi\times\text{6.8}$GHz for $|g\rangle = |5S_{1/2}, F= 2\rangle$, $|s\rangle = |5S_{1/2}, F=1\rangle$ and $|e\rangle = |5P_{1/2}, F' = 2\rangle$ as reported in \cite{Pu2017}. The angular wave-function of the idler photon is calculated by sampling the $\theta_{k},\phi_{k}$ dependent part of Eq.~(\ref{eq:tranread}) (without the $g_{eg,\tau}$ term, which is taken to be a constant according to the argument below Eq.~(\ref{eq:idlerfreq})) and is normalized numerically. Then we calculate its overlap with the normalized
Gaussian mode of Eq.~(\ref{eq:idler}) to get the IRE $\eta$.\\

\begin{figure}
  \includegraphics[width=\linewidth]{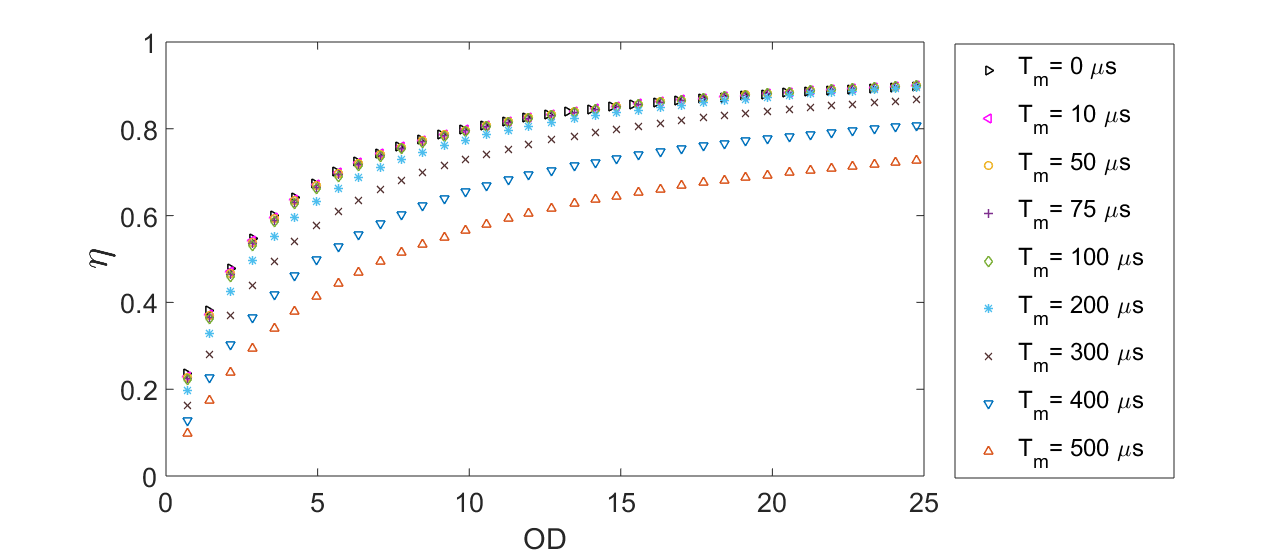}
  \caption{The intrinsic retrieval efficiency as a function of the optical depth for increasing memory storage times $T_{m}$ with skew angle fixed to be 0.}
  \label{fig:Tmangle0}
\end{figure}

First, we will look at the ideal case of stationary atoms, implying a storage
time $T_{m} = 0$. The IRE thus evaluated is independent of storage time. In
Fig.~(\ref{fig:Tmzero}), we observe that $\eta$ always remains smaller than
unity for the given optical depth OD = 24.7, and different values of skew
angle, $\Theta$, as a function of the width ratio (WR) between the write-pulse
and the optical fiber mode waists:
\begin{figure}
  \includegraphics[width=\linewidth]{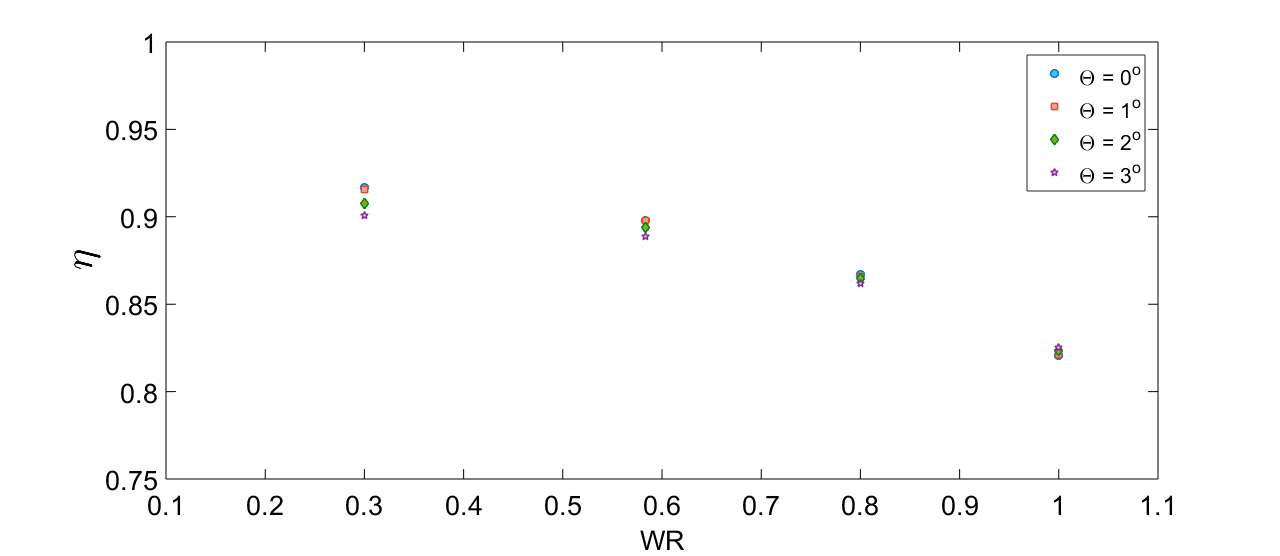}
  \caption{Intrinsic retrieval efficiency $\eta$ as a function of the width ratio WR between the waist width of the signal (idler) optical fiber mode over that of the write beam for different values of skew angle.}
  \label{fig:Tmzero}
\end{figure}
\begin{eqnarray}
WR &=& \frac{W_{i}}{W_{w}} = \frac{W_{s}}{W_{w}}
\end{eqnarray}
As we can see, $\eta$ increases with decreasing WR. The reason $\eta$
cannot reach 1 is that there is a mismatch between the photon profile and the
optical fiber mode. Fig (\ref{fig:Tmangle0}) captures the variation of the IRE as a function of
the optical depth of the system for different values of $T_{m}$ with
$\Theta=0^{o}$ and WR = 35$\mu$m/60$\mu$m fixed. 
The OD is adjusted by changing the atomic density while keeping
the beam parameters constant. \\

\begin{figure}
\centering
    \subfloat[Re\text{[} $f^{r}(\theta_{k},\phi_{k})$\text{]}]{\label{fig:77Real}\includegraphics[width=0.25\textwidth]{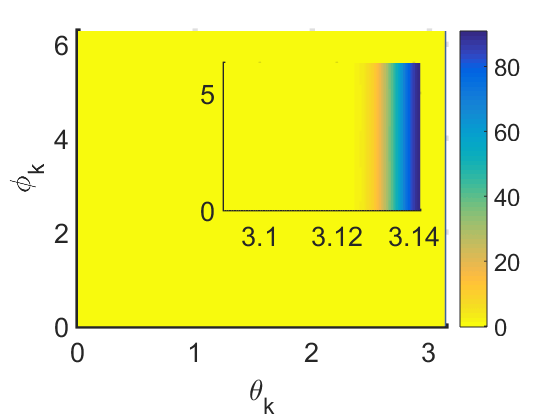}}~              
    \subfloat[Im\text{[}$f^{r}(\theta_{k},\phi_{k})$\text{]}]{\label{fig:77Imag}\includegraphics[width=0.25\textwidth]{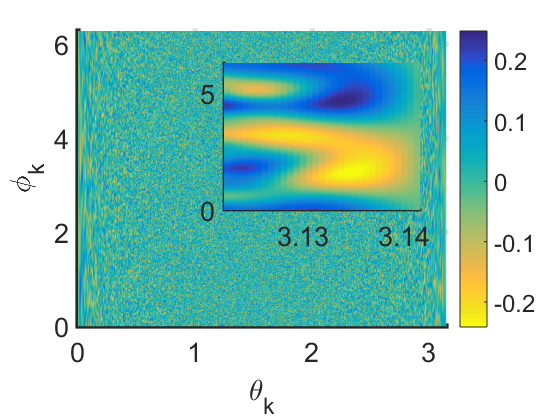}}
\caption{The normalized angular mode function, $f^{r}(\theta_{k},\phi_{k})$, at OD = 24.7, $\Theta = 0^{o}$, WR = 35/60 and $\text{T}_{m} =$ $0\,\mu$s.}\label{fig:77}
\end{figure}

We see the signature of collective enhancement
as has been proved in \cite{Duan02}. The output photon mode that is correlated
with the atomic spin wave has higher fractional contribution along the $\theta_{k}=\pi$ direction which increases as the number of atoms goes up. The normalized angular mode $\hat{f}%
^{r}(\theta_{k},\phi_{k})$ for the idler photon obtained for a dense atomic ensemble is shown in Fig.~(\ref{fig:77}) for $T_{m} = 0$ and $\Theta = 0^{o}$. This angular profile for an atomic sample with OD = 24.7 and for WR =
35/60 gives about 90\% IRE. 


The real part of the angular mode profile, in the absence of
decoherence effects due to non-zero $T_{m}$ and $\Theta$, is plotted in
Fig.~(\ref{fig:77Real}). It clearly shows a pronounced emission peak near
angle $\theta_{k} = \pi$ (shown in the inset) for all azimuthal angles. Apart from the emission around the $\theta_{k} = \pi$ direction, there are noisy contributions present along all other directions as well. The idler photon mode profile has contributions that are prominently from the real part as expected. Without any atomic density fluctuations, that is, replacing the summation over atoms in Eq.~(\ref{eq:tranread}) with a continuous integration, the imaginary part of the mode function would be identically zero. Thus,
imaginary part of the angular profile gives us a scale of fluctuations in all
the directions. These fluctuations are related to the density fluctuations of
the atomic sample. Important feature to note is that the scale of these fluctuations
is very small compared to the scale of the enhanced photon emission to be collected.
It is a function of OD and $\Theta$; with decreasing OD and increasing skew
angle, we see the relative contributions of the fluctuations in all directions
go up.\\  

There is a limit to increasing the optical depth by raising the atomic density
because the low atomic density assumption would then breakdown and effects of
atom-atom interactions mediated by light will have to be considered
\cite{Dicke54}.\\

\begin{figure}
  \includegraphics[width=\linewidth]{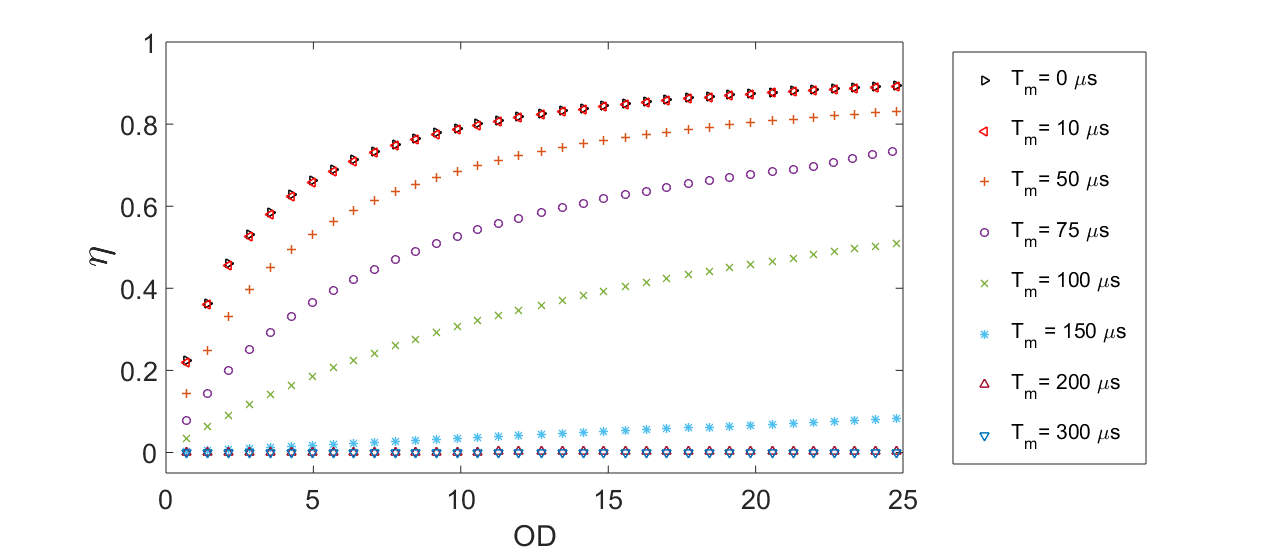}
  \caption{The intrinsic retrieval efficiency as a function of optical depth for increasing storage times $T_{m}$ with skew angle $\Theta=2^{o}$}
  \label{fig:Tmangle2}
\end{figure}

\begin{figure}
  \includegraphics[width=\linewidth]{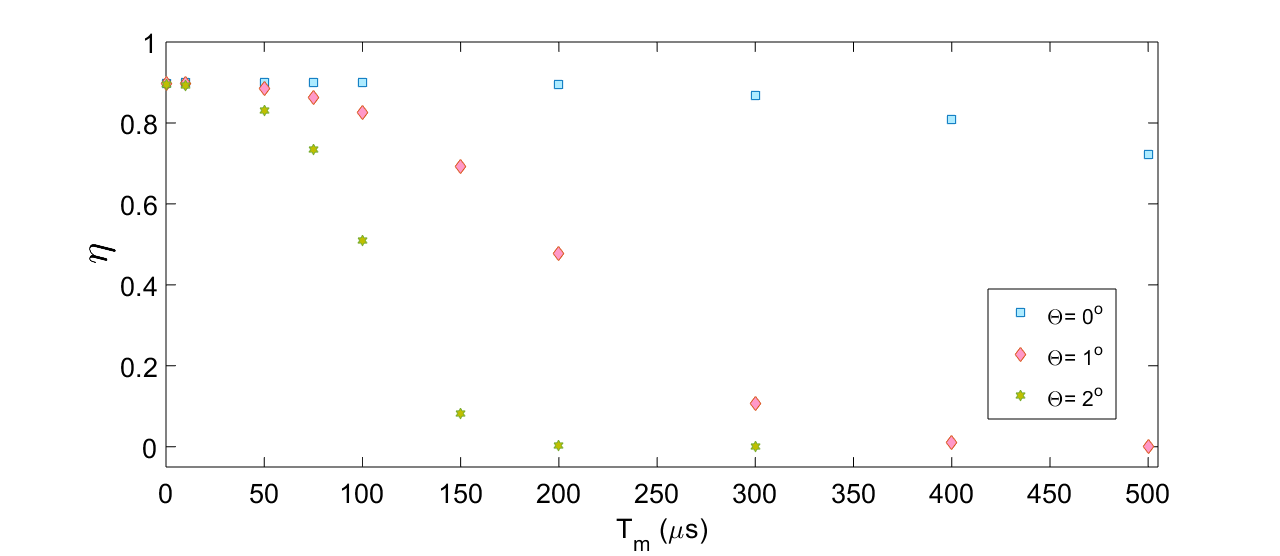}
  \caption{At optical depth = 24.7, intrinsic retrieval efficiency varies as a function of the memory storage time $T_{m}$ for skew angle values $\Theta= (0^{o},1^{o},2^{o})$}
  \label{fig:Tmallang}
\end{figure}

Now let us look at the effect of non-zero $T_{m}$ values for skew angle
$\Theta= 2^{o}$ and WR = 35$\mu$m/60$\mu$m which correspond to the
experimental value of parameters from the Tsinghua setup \cite{Pu2017}.
Fig.~(\ref{fig:Tmangle2}) shows the variation of the IRE as a function of OD
for different values of $T_{m}$ at $\Theta=2^{o}$. 
Comparing Fig.~($\ref{fig:Tmangle0}$) for $\Theta= 0^{o}$ and
Fig.~($\ref{fig:Tmangle2}$) for $\Theta= 2^{o}$, we see the effect of
decoherence due to misalignment between the write-read and the signal-idler
electric fields. The IRE falls from 80\% for $T_{m} = 0\,\mu s$ to 50 \% for
$T_{m} =100\,\mu s$ when skew angle is $2^{o}$ for OD of 24.7 compared to no
noticeable change in the $\eta$ value (90\%) for $T_{m}$ increasing from $0$
to $100\,\mu s$ when skew angle is set to $0^{o}$. The variation in the IRE
for different skew angles and memory storage times at a fixed OD = 24.7 are
shown in Fig.~(\ref{fig:Tmallang}).
We see a rapid decrease in the IRE for non-zero skew angles as the
memory storage time is increased. For a retrieval efficiency larger than 80\%
we can store the atomic spin wave for a maximum of 50 $\mu s$ with $\Theta=
2^{o} $ which is not sufficient for implementation of DLCZ quantum repeater
protocol efficiently. An important point that must be mentioned here is that
the IRE can be increased by using optical traps for the atomic ensemble which
restrict the atomic motion and hence help reduce atomic motion induced
decoherence, though even after the implementation of such traps, it is still
not possible to reach unit retrieval efficiency. Our current theoretical model
can be extended to include the effects of optical traps by changing the
expression for the atomic positions in Eq.~(\ref{eq:newpos}) appropriately.\\

\begin{figure}
\centering
    \subfloat[Re\text{[}$f^{r}(\theta_{k},\phi_{k})$\text{]}]{\label{fig:112Real}\includegraphics[width=0.25\textwidth]{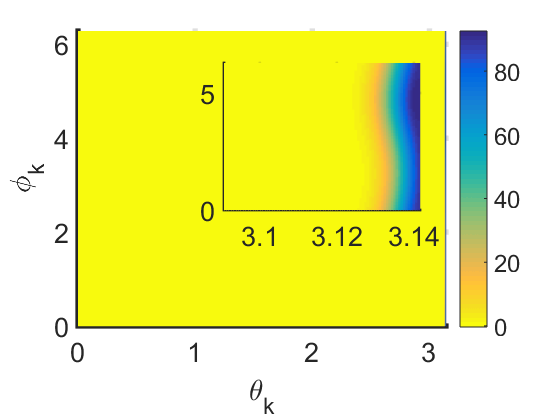}}~      
    \subfloat[Im\text{[}$f^{r}(\theta_{k},\phi_{k})$\text{]}]{\label{fig:112Imag}\includegraphics[width=0.25\textwidth]{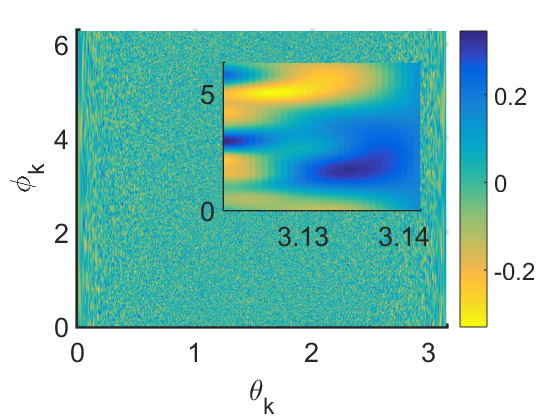}}
\caption{The normalized angular mode function, $f^{r}(\theta_{k},\phi_{k})$, at OD = 24.7, $\Theta = 1^{o}$, WR = 35/60 and $T_{m} = 100\,\mu s$.}\label{fig:112}
\end{figure}

\begin{figure}[h]
\centering
    \subfloat[Re\text{[}$f^{r}(\theta_{k},\phi_{k})$\text{]}]{\label{fig:120Real}\includegraphics[width=0.25\textwidth]{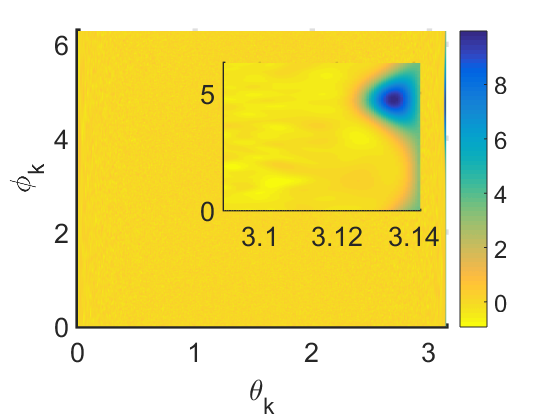}}~   
    \subfloat[Im\text{[}$f^{r}(\theta_{k},\phi_{k})$\text{]}]{\label{fig:120Imag}\includegraphics[width=0.25\textwidth]{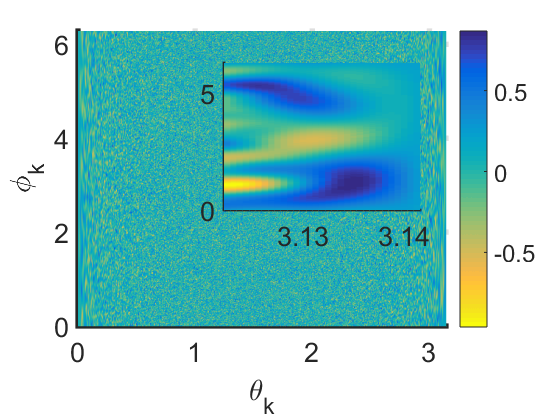}}
\caption{The normalized angular mode function, $f^{r}(\theta_{k},\phi_{k})$, at OD = 24.7, $\Theta = 2^{o}$, WR = 35/60 and $T_{m} = 200\,\mu s$.}\label{fig:120}
\end{figure}

Let us also look at the angular profile for non-zero skew angles and memory
storage times. Specifically, we choose a configuration of parameters that
gives around $\eta$ = 80\%, particularly, $\Theta= 1^{o}$ and $T_{m} =
100\,\mu s$ [Fig.~({\ref{fig:112}})] and compare it with a value of $\eta$ =
0.3\% for $\Theta= 2^{o}$ and $T_{m} = 200\,\mu s$ [Fig.~(\ref{fig:120})].

We see that Fig.~(\ref{fig:112Real}) shows a prominent
contribution around $\theta_{k} = \pi$. On close observation, as shown in the
inset, we can detect slight variation in the transverse profile along the
$\phi_{k}$ direction for $\theta_{k} \approx\pi$, which becomes more
pronounced with larger skew angle and longer storage time in
Fig.~(\ref{fig:120Real}). The $\theta_{k}$ and $\phi_{k}$ dependence of the
observed mode profiles can be attributed to the disruption of symmetry in the
z-direction due to non-zero skew angle. As already mentioned, the imaginary
part of the mode profile gives an insight about the fluctuations present in
all the directions that do not have overlap with the optical fiber electric
field. These fluctuations are present in the real part as well, but get washed
out by the dominant contribution of the idler photon. Fluctuations in the mode
profile are also caused by the atomic density fluctuations in the sample. The
fluctuations observed in Fig.~(\ref{fig:112Imag}) are of the same order as those observed in Fig.~(\ref{fig:77Imag}). In Fig.~(\ref{fig:120Real}) we see
higher contribution to the mode profile from all values of $\theta_{k}$ and
$\phi_{k}$ when compared to Fig.~(\ref{fig:77Real}) and
Fig.~(\ref{fig:112Real}), and the fluctuations are significantly higher as
seen from Fig.~(\ref{fig:120Imag}). With this we conclude the discussion of
the numerical results.
\section{\label{sec:level7}Discussion}
We have formulated a three-dimensional theory to study the intrinsic retrieval
efficiency (IRE) during the write-read process for quantum repeater protocols.
The focus of this calculation was to describe the quantum mechanical process
involved in the interaction of the atomic ensemble with the control light
pulses in a three-level $\Lambda$ system. The motivation for this work was
primarily to understand the factors that influence the IRE which plays a
crucial role in the success of quantum repeater protocols like DLCZ method and
its variants \cite{Sangouard11}. \newline

Different interaction strengths involved in the write process and read process
were looked at separately. The quantum state obtained by perturbative analysis
in the write process provides us with the initial condition for the quantum
evolution during the read process. An important result obtained from this
calculation is the expression of the IRE as a function of the parameters of
the atomic ensemble and control pulses. We show that unit retrieval efficiency
is not possible for realistic experimental parameters. \newline

We also show the effects of decoherence introduced due to atomic motion in the
sample, which drastically reduce $\eta$ for the skewed configuration of atomic
beams. Neglecting the atomic motion for the duration of write and read pulses,
within which the accumulated phase is small, only the change in atomic
positions during the storage period contributes to the decoherence. In
general, for ballistic motion of atoms in the absence of collisions, the
average separation between atoms increases with time and the IRE decreases.
This can be corrected by using atomic traps which limit the atomic motion. On
average the atomic separations with increasing storage times are constant in
atomic traps thus improving the atomic retrieval efficiency immensely
(\cite{RZhao09},\cite{Yang16}). \newline

\bibliography{research}

\end{document}